\documentclass[aps, prb, notitlepage, superscriptaddress, longbibliography, twocolumn, 10pt]{revtex4-2}
\usepackage{xcolor}
\usepackage{amsmath, amsthm, amssymb, bbold}

\usepackage{graphicx}
\usepackage[normalem]{ulem}
\usepackage{cancel}
\usepackage{multirow}

\usepackage{bm}

\usepackage{microtype}
\usepackage{hyperref}

\hypersetup{
    colorlinks=true,
    linkcolor=blue,
    filecolor=magenta,      
    urlcolor=blue,
}

\newcommand{\hc}{{\rm H.c.}}

\newcommand{\avg}[1]{\langle #1\rangle}
\newcommand{\mc}[1]{\mathcal{#1}}

\newcommand{\zavg}[1]{\langle\!\langle #1\rangle\!\rangle}

\def\im{\text{Im}}
\def\tr{\text{Tr}}
\def\sgn{\mathrm{sgn}}

\def\id{\mathbb{1}}
\def\llangle{\langle\negmedspace\langle}
\def\rrangle{\rangle\negmedspace\rangle}

\begin{document}
\title{Interaction-induced asymmetry in infinite-temperature dynamical correlations of hard-core anyons}

\author{Doru Sticlet}
\email{doru.sticlet@itim-cj.ro}
\affiliation{National Institute for R\&D of Isotopic and Molecular Technologies, 67-103 Donat, 400293 Cluj-Napoca, Romania}
\author{Ovidiu I. P\^ a\c tu}
\affiliation{Institute for Space Sciences, Bucharest-M\u agurele, R 077125, Romania}
\author{Bal\'azs D\'ora}
\affiliation{Department of Theoretical Physics, Institute of Physics, Budapest University of Technology and Economics, M\H{u}egyetem rkp.~3, H-1111 Budapest, Hungary}
\affiliation{MTA-BME Lend\"ulet ``Momentum'' Open Quantum Systems Research Group, Institute of Physics, Budapest University of Technology and Economics,
M\H{u}egyetem rkp.~3, H-1111 Budapest, Hungary}
\author{C\u at\u alin Pa\c scu Moca}
\email{mocap@uoradea.ro}
\affiliation{Department of Theoretical Physics, Institute of Physics, Budapest University of Technology and Economics, M\H{u}egyetem rkp.~3, H-1111 Budapest, Hungary}
\affiliation{MTA-BME Lend\"ulet "Momentum" Open Quantum Systems Research Group, Institute of Physics, Budapest University of Technology and Economics,
M\H{u}egyetem rkp. 3., H-1111 Budapest, Hungary}
\affiliation{Department of Physics, University of Oradea,  410087 Oradea, Romania}

\begin{abstract}
We study dynamical correlations of interacting hard-core anyons on a one-dimensional lattice at infinite temperature.
This is a setting in which the many-body spectrum is independent of the statistical phase $\theta$, while dynamical correlators remain sensitive to $\theta$ through nonlocal Jordan-Wigner strings.
We compute single-particle Green's functions, spectral functions, and density-density correlators, thereby separating the effects of fractional statistics on one-body coherence from those on density transport in a maximally mixed ensemble.
In the noninteracting case $V=0$, high-temperature averaging leads to inversion-symmetric Green's functions for all $\theta$ despite the presence of anyonic strings.
Finite nearest-neighbor interactions $V$ generate, however, a pronounced left-right asymmetry in the Green's functions for $0<\theta<\pi$, with the strongest chirality 
appearing at intermediate couplings $V\sim J$ where interactions and hopping compete most effectively. In this regime, the Green's function decays exponentially in time with
a statistical-angle-dependent decay rate.
At strong coupling, the dynamics crosses over to an atomic-limit regime in which the dependence on $\theta$ is reduced. 
Here, the Green's function decays universally as $t^{-1}$ and the corresponding spectral function displays a three-band structure.
In contrast, density-density correlations are insensitive to statistics and recover the known infinite-temperature transport regimes of the XXZ chain, including ballistic, superdiffusive, and diffusive behaviors.
These results identify dynamical correlation functions as direct probes of fractional statistics in high-entropy quantum systems.
\end{abstract}
	
\maketitle
\section{Introduction}

Anyons were originally introduced as quasiparticles in two dimensions whose wave function acquires a fractional phase upon exchange~\cite{Leinaas1977}.
The notion of fractional statistics was later generalized beyond this setting through the generalized Pauli exclusion principle~\cite{Haldane1991}, and it now plays a central role in topics ranging from fractional quantum Hall physics~\cite{Arovas1984,Moore1991} and spin liquids~\cite{Kalmeyer1987,Kitaev2006} to topological quantum computation~\cite{Nayak2008}.

One-dimensional anyons have attracted sustained interest since the first studies of continuum models such as the Tonks-Girardeau and Lieb-Liniger gases~\cite{Kundu1999,Girardeau2006,Batchelor2006,Patu2007,Calabrese2007}.
On the lattice, two broad classes of anyonic models have been studied most extensively.
The first allows multiple occupancy of a site, so that the particles behave as bosons locally while acquiring a fractional phase upon exchange between different sites.
This setting underlies the anyon Bose-Hubbard model~\cite{Kundu2010,Keilmann2011,Greschner2015,Tang2015,Straeter2016,ArcilaForero2016a,Lange2017,Liu2018,Zuo2018,Bonkhoff2021}, where on-site and nearest-neighbor interactions generate a rich phase diagram including superfluid, Mott-insulating, charge-density-wave, and dimerized phases~\cite{Bonkhoff2025}.
Recent experimental progress toward implementing anyon models in cold-atom and synthetic quantum systems has further strengthened the motivation for such studies~\cite{Kwan2024,Dhar2025}.

Here, we focus on the second class, namely, hard-core anyons.
For single-flavor anyons, the local Hilbert space is two dimensional, so the particles behave as spinless fermions on site, while exchanges on different sites still acquire a nontrivial statistical phase.
This makes hard-core anyons especially appealing: The Hamiltonian spectrum is insensitive to the statistical angle, whereas dynamical correlation functions remain sensitive to it through nonlocal Jordan-Wigner strings.
This property has motivated extensive theoretical research on hard-core anyons. 
Early studies focused primarily on ground-state properties, including persistent currents~\cite{Zhu1996}, one-dimensional spin models formulated in terms of hard-core anyonic statistics~\cite{Amico1998}, zero-temperature phases of interacting anyons~\cite{Mila2008}, and the asymmetric momentum distribution characteristic of anyonic systems~\cite{Hao2009}. 
More recently, the steady state and the Liouvillian gap was studied in the context of dephasing dynamics in a hard-core anyon chain~\cite{,Guan2025}. On the experimental side, hard-core anyons could be realized with ultracold atoms as a limiting case of previous proposals to realize the anyon Bose-Hubbard model in optical lattices~\cite{Greschner2015,Greschner2018}.

Dynamical aspects subsequently attracted considerable attention, particularly in the contexts of quantum quenches~\cite{Campo2008,Hao2012,Wright2014} and quantum walks~\cite{Wang2014}. 
These works largely focused on equal-time correlations and expansion dynamics with emphasis on quantum-quench relaxation of one-body observables, such as the single-particle density matrix and related equal-time quantities.
Hard-core anyons were also predicted to produce experimentally accessible transport signatures in quasi-one-dimensional antidot geometries~\cite{Averin2007}.
In contrast, unequal-time correlation functions have received comparatively less attention and have typically been studied only in the absence of interactions. 
Such correlation functions of impenetrable anyons have been investigated both in the continuum~\cite{Patu2007} and on the lattice~\cite{Patu2015,Wang2022}, at zero and finite temperature. 
The latter work~\cite{Wang2022}, in particular, observed that high temperatures wash out the inversion asymmetry for noninteracting anyons and highlighted the need for further investigation into the role of interactions.

In this work, we study the infinite-temperature limit, where all many-body states are equally probable.
This regime is both conceptually clean and experimentally relevant.
It removes complications associated with low-temperature order, symmetry breaking, and the detailed role of conserved quantum numbers, while remaining directly relevant to quantum-simulation experiments that prepare high-energy or randomly populated states and then probe their ensuing dynamics~\cite{Nahum2017,Fisher2023}.
At the same time, recent work has shown that infinite-temperature dynamics can still display highly nontrivial operator spreading and anomalous transport~\cite{Ljubotina2017,Ljubotina2019,Valli2025,Moca2025}.
This makes it a natural setting in which to isolate the interplay between interactions and fractional statistics.

We address two related questions.
How visible is anyonic exchange statistics in real-time correlation functions of a maximally mixed ensemble, and how do interactions reshape these correlations, in particular, their spreading, decay, and spectral structure?

To answer these questions, we analyze several real-time observables at $T=\infty$.
Specifically, we compute single-particle Green's functions, spectral functions, and density-density correlators.
Our main findings are as follows.
For $V=0$, the infinite-temperature Green's functions are inversion symmetric for all statistical phases $\theta$.
Once nearest-neighbor interactions are switched on, however, a pronounced left-right asymmetry emerges for anyons with $0<\theta<\pi$.
This chirality is strongest at intermediate couplings $V\sim J$, whereas at strong interactions the dynamics crosses over toward an atomic-limit regime in which the dependence on statistics is progressively reduced.
In this regime, the Green's functions acquire a universal time decay $\sim t^{-1}$ independent of the statistical phase.

Our study is also directly relevant to the XXZ spin chain at infinite temperature, which has become a canonical setting for investigating anomalous transport~\cite{Znidaric2011,Ljubotina2017,Ljubotina2019,Gopalakrishnan2023}.
In the bosonic limit, the anyon model maps to the XXZ chain, and the retarded Green's functions map to transverse spin-spin correlators.
This allows us to connect the anyonic problem to the established XXZ transport phenomenology and to compare the transverse and longitudinal dynamical sectors within a common framework.

The article is organized as follows. 
Section~\ref{sec:model} introduces the anyon model, the operator representations used throughout the paper, and its relation to the XXZ spin chain. 
Section~\ref{sec:correlations} introduces the infinite-temperature ensemble, defines the correlation functions and spectral quantities of interest, and summarizes the symmetry constraints used later in the discussion.
Section~\ref{sec:results} presents the main results for the Green's functions at zero and finite interaction strength, followed by the spectral functions and density-density correlations.
Finally, Sec.~\ref{sec:conclusions} summarizes the main conclusions of the present work.

\section{Interacting hard-core anyons}\label{sec:model}
We consider the following Hamiltonian for hard-core anyons on a one-dimensional lattice of $L$ sites~\cite{Amico1998,Hao2009}:
\begin{equation}\label{eq:ham}
H = \sum_{j=1}^{L-1}J(a_j^\dag a_{j+1} + \hc) 
+ V \left(n_j-\frac12\right) \left(n_{j+1} -\frac12\right),
\end{equation}
with anyonic creation and annihilation operators $a_j^\dag$ and $a_j$ at site $j$, and the local density operator $n_j = a_j^\dag a_j$.
The hopping amplitude $J$ sets the energy scale and is taken to be positive. 
Throughout the article, we work in units where $\hbar$, the Boltzmann constant $k_B$, and the lattice constant are set to 1.
Without loss of generality, the nearest-neighbor interactions are chosen in simulations to be non-negative, $V\geq0$. 
As we show in Sec.~\ref{sec:symmetry}, there is a symmetry that relates the correlations at positive and negative $V$, such that
the sign of $V$ does not affect the main conclusions of the present work.

Throughout this work, we use three equivalent operator languages to describe the local Hilbert space for hard-core particles: anyonic operators $a_j$, hard-core bosonic operators $b_j$, and fermionic operators $c_j$.
The anyonic operators are related to the bosonic and fermionic ones through the generalized Jordan-Wigner transformations~\cite{Girardeau2006,Hao2009,Hao2012},
\begin{equation}\label{eq:JW}
a_j = b_j e^{i\theta \sum_{k<j} n_k}= c_j e^{i(\theta-\pi) \sum_{k<j} n_k}, 
\end{equation}
with $\theta$ the statistical phase, which interpolates between bosons ($\theta=0$) and fermions ($\theta=\pi$).
The local density operator is identical for all three languages, $n_j = a_j^\dag a_j = b_j^\dag b_j = c_j^\dag c_j$, due to cancellation of the Jordan-Wigner strings.

The anyon operators satisfy the following commutation relations:
\begin{align}\label{eq:comm}
a_j a_k^\dag &+ e^{-i(\pi-\theta) \sgn(j-k)} a_k^\dag a_j = \delta_{jk},\notag\\
a_j a_k &+ e^{i(\pi-\theta) \sgn(j-k)} a_k a_j = 0,
\end{align}
with $\sgn(x)=x/|x|$ for $x\neq0$ and $\sgn(0)=0$.
These expressions ensure the hard-core constraints on the same site, $a_j^2=0$ and $\{a_j,a^\dag_j\}=1$.

When Eq.~\eqref{eq:ham} is rewritten in the bosonic or fermionic language, the Jordan-Wigner strings between nearest-neighbor hopping operators cancel out when considering the action of $H$ on the Fock space.
As a result, the Hamiltonian has no explicit dependence on the statistical phase $\theta$, and the energy spectrum is independent of $\theta$.

In contrast to the Hamiltonian, the dynamical correlation functions, which are the main focus of the present study, are sensitive to the anyon statistics.
Due to the twisted Jordan-Wigner transformation~\eqref{eq:JW}, the correlations involve nonlocal string operators that render the dynamics nontrivial.
The infinite-temperature limit is particularly interesting because it defies the conventional wisdom that the effect of statistics is washed out by thermal fluctuations.
Instead, we find that interactions induce strong spatial asymmetries in the anyonic dynamical correlations.

\emph{The bosonic limit and relation to the XXZ spin chain.} Our study is also relevant for recent interest in the XXZ spin chain at infinite temperature, where anomalous transport and operator spreading have been intensively studied~\cite{Ljubotina2017,Ljubotina2019,Gopalakrishnan2019a,DeNardis2019,Gopalakrishnan2019,Dupont2020,Bulchandani2021,Gopalakrishnan2023}.
The connection becomes explicit in the bosonic representation.
Hard-core bosons are identified with spin-$1/2$ operators as
\begin{equation}
b_j^\dag = S_j^-,\quad b_j = S_j^+,\quad n_j = \frac12(\mathbb{1} - \sigma_j^z),
\end{equation}
where $S_j^{\pm}=(\sigma_j^x \pm i\sigma_j^y)/2$ and $\sigma_j^{x,y,z}$ are the Pauli matrices at site $j$.
Using these relations in Eq.~\eqref{eq:ham}, one obtains the XXZ Hamiltonian
\begin{equation}
H = \sum_{j=1}^{L-1} \frac{J}{2}(\sigma_j^x \sigma_{j+1}^x + \sigma_j^y \sigma_{j+1}^y) + \frac{V}{4} \sigma_j^z \sigma_{j+1}^z,
\end{equation}
with $V$ playing the role of the spin anisotropy.
In this language, bosonic Green's functions map to transverse spin-spin correlators, while density-density correlators map to longitudinal spin correlations.
At infinite temperature, the system is known to display ballistic transport for $V<2J$, superdiffusive transport [in the Kardar-Parisi-Zhang (KPZ) universality class~\cite{Kardar1986}] at the isotropic point $V=2J$, and diffusive transport for $V>2J$ for non-negative $V$ and $J$.

\section{Correlation functions at infinite temperature}\label{sec:correlations}
We now introduce the dynamical observables studied in the infinite-temperature ensemble.
In the $T=\infty$ limit, all many-body states are equally probable and the density matrix becomes proportional to the identity, $\rho_\infty=\id/\mathcal D$, with $\mathcal D$ the size of the Hilbert space; that is, $\mathcal D = 2^L$ for the model~\eqref{eq:ham}.
Accordingly, the expectation value of an operator (or product of operators) $\Pi(t)$ is
\begin{equation}
\langle \Pi(t) \rangle = \frac{\tr[\rho_\infty\,\Pi(t)]}{\tr[\rho_\infty]},\quad  \langle \Pi(t) \rangle=\frac{1}{\mathcal D}\tr[\Pi(t)].
\label{eq:expectation}
\end{equation}

An immediate consequence is that equal-time one-body correlations are trivial,
\begin{equation}
\avg{a_j(t) a_k^\dag(t)} = \frac12 \delta_{jk},
\label{eq:equal_time}
\end{equation}
since the density matrix commutes with all operators.
Therefore, the local density is uniform at all times, $\avg{n_j(t)} = \frac12$.
This is in stark contrast to the finite-temperature case, where the interplay between statistics and interactions can lead to nontrivial spatial asymmetries in the local density evolution for the anyon Bose-Hubbard model~\cite{Liu2018}.

In the following, we define the single-particle Green's functions, the spectral function, and the density-density correlators used throughout the paper.

\subsection{Green's functions}\label{sec:green_definitions}
Since equal-time correlations are trivial at infinite temperature, the simplest nontrivial correlators that show the effect of anyonic statistics are the dynamical one-body Green's functions.
These involve correlators of the form $\avg{a_j(t)a^\dag_k}$ that have a nontrivial dependence on the statistical phase $\theta$ through the Jordan-Wigner strings.
For example, by using the generalized Jordan-Wigner transformation~\eqref{eq:JW}, and mapping anyons to hard-core bosons represented by operators $b_j$ and $b_j^\dag$, one finds
\begin{equation}
\avg{a_j(t)a^\dag_k} 
= \avg{
e^{iHt} b_je^{i\theta \sum_{k'<j} n_{k'}} e^{-iHt} e^{-i\theta\sum_{k'<k} n_{k'}} b_k^\dag
}.
\end{equation}
Similar expressions hold for the mapping to fermions, such that the above  anyonic correlator can be viewed as a string-dressed one-body correlator of either fermions or hard-core bosons.

The retarded Green's function is central to understanding the response of the system to the addition or removal of a particle. 
Thus, the retarded Green's function represents a natural probe of fractional statistics since adding or removing an anyon at a given site is sensitive to the presence of other anyons on the lattice through the nontrivial commutation relations with them resulting from the nonlocal nature of Jordan-Wigner strings.
In the context of anyons implemented in ultracold atoms in one-dimensional optical lattice, the retarded Green's function can in principle be accessed using many-body Ramsey interferometry~\cite{Knap2013}.
The retarded Green's function for anyons reads
\begin{equation}\label{eq:Gtjk}
G^{R}_{jk}(t;\theta) = -i \Theta(t) \langle [a_j(t), a_k^\dag(0)]_\theta \rangle, 
\end{equation}
with $\Theta(t)$ the Heaviside step function, and where the $\theta$ commutator $[\cdot,\cdot]_\theta$ is defined such that in the limit $t\to 0^+$ the commutation relation~\eqref{eq:comm} of anyon operators is recovered.
Explicitly, the retarded Green's function depends on the statistical phase $\theta$ as
\begin{equation}
G^{R}_{jk}(t;\theta) = -i\Theta(t)\langle a_j(t) a_k^\dag + 
e^{-i(\pi-\theta) \sgn(j-k)} a_k^\dag a_j(t)
\rangle,
\end{equation}
where the anyon operators without argument are evaluated at time $t=0$.
This candidate Green's function reduces to the conventional bosonic and fermionic retarded Green's functions in the limits $\theta=0$ and $\theta=\pi$, respectively, when the sites $j$ and $k$ are distinct.
At equal sites $j=k$, the anyon operators obey hard-core constraints, and therefore one recovers the fermionic equal-site Green's functions,
\begin{equation}
G^{R}_{jj}(t;\theta) = -i\Theta(t) \langle \{a_j(t), a_j^\dag\} \rangle,
\end{equation}
with $\{\cdot,\cdot\}$, the anticommutator.
The retarded Green's function is expressed in terms of the greater and lesser Green's functions,
\begin{equation}
G^{R}_{jk}(t;\theta) = \Theta(t) [G^{>}_{jk}(t) - G^{<}_{jk}(t)],
\end{equation}
and one identifies the greater Green's function
\begin{equation}
G^{>}_{jk}(t;\theta) = -i\langle a_j(t) a_k^\dag \rangle,
\end{equation}
and the lesser Green's function
\begin{equation}
G^{<}_{jk}(t;\theta) = ie^{i(\theta-\pi)\sgn(j-k)}\langle a_k^\dag a_j(t) \rangle.
\end{equation}
As expected, these choices are again consistent with the conventional Green's functions in the bosonic and fermionic limits. 

At infinite temperature, the greater and lesser Green's functions are related by
\begin{equation}\label{eq:lesserTinf}
G^<_{jk}(t;\theta) = - e^{i(\theta-\pi)\sgn(j-k)} G^{>}_{jk}(t;\theta),
\end{equation}
using the cyclic property of the trace.
Henceforth, we will drop the dependence on $\theta$ in the Green's function arguments and related correlators, unless the $\theta$ dependence is explicitly needed to describe a specific statistical phase.

\subsection{Bosonic limit and XXZ spin correlators}
\label{sec:xxz_correlations}
In the bosonic limit $\theta=0$, the anyon operators reduce to hard-core bosons, $a_j=b_j$ and $a_j^\dag=b_j^\dag$.
Using the XXZ mapping introduced in Sec.~\ref{sec:model}, one obtains the explicit correspondence
\begin{align}
G^>_{jk}(t;\theta=0) &= -i\avg{S_j^+(t) S_k^-(0)},\notag\\
G^<_{jk}(t;\theta=0) &= i\avg{S_k^-(0) S_j^+(t)},
\end{align}
showing that the bosonic Green's functions probe transverse spin dynamics in the XXZ chain.
For the local greater Green's function used repeatedly below, this reduces to
\begin{equation}\label{eq:transverse}
G^>(0,t;\theta=0) = -i\avg{S_0^+(t) S_0^-(0)}.
\end{equation}
Thus, in the bosonic limit, the Green's functions access the transverse spin sector of the XXZ chain. Throughout this work, however, we present the results in terms of creation and annihilation operators and use the spin-language correspondence only as an interpretation of the bosonic limit.

\subsection{Spectral functions and density of states}\label{sec:spectral}
The spectral function provides another direct probe of fractional statistics which can in principle be experimentally probed using momentum-resolved rf and Raman spectroscopy~\cite{Bohrdt2017,Vale2021}.
The spectral function is obtained from the Fourier transform of the retarded Green's function,
\begin{equation}\label{eq:A_q_omega}
A(q,\omega) = -\frac{1}{\pi} \im \int dt\sum_x e^{i(\omega t-qx)} G^R(x,t),
\end{equation}
where $k=0$ denotes the reference site and the second spatial index is written as $j\to x$.
The local density of states 
\begin{equation}
\rho(\omega)\equiv A(x=0,\omega) = -\frac{1}{\pi}\im\, G^R(x=0,\omega)
\end{equation}
follows from the previous Eqs.~\eqref{eq:lesserTinf} and \eqref{eq:A_q_omega} as
\begin{align}\label{eq:dos}
\rho(\omega)
&= -\frac{2}{\pi} \im \int_0^\infty dt e^{i\omega t} G^>(x=0,t).
\end{align}
In the numerical calculations, the greater Green's function $G^>(x,t)$ is computed first, from which the retarded Green's function and the spectral observables are obtained.

\subsection{Density-density correlators}\label{sec:density_density}
The density-density correlators and their connected versions are denoted by
\begin{align}
C_{jk}(t) &= \avg{n_j(t)n_k(0)},\nonumber\\
C^c_{jk}(t) &= C_{jk}(t) - \avg{n_j(t)}\avg{n_k(0)}.
\end{align}
Unlike the Green's functions, they probe the dynamics of the conserved density rather than the propagation of a single anyonic operator.
Since the number operators do not depend on the anyonic strings, these correlators have no explicit dependence on the statistical phase $\theta$.
In the bosonic limit $\theta=0$, they map to the longitudinal spin sector of the XXZ chain introduced in Sec.~\ref{sec:model}.
Since $n_j=1/2-S_j^z$, with $S_j^z=\sigma_j^z/2$, one has
\begin{equation}
C_{jk}(t) = \avg{S_j^z(t)S_k^z(0)} + \frac14,
\quad
C^c_{jk}(t) = \avg{S_j^z(t)S_k^z(0)},
\end{equation}
where the first identity follows from $\avg{S_j^z}=0$ at infinite temperature.
Thus the connected density-density correlator directly probes longitudinal spin correlations, complementary to the transverse correlations encoded in the bosonic Green's functions discussed in Sec.~\ref{sec:xxz_correlations}.
Nevertheless, they provide direct information about density fluctuations and about the transport regime selected by the interaction strength $V$.

\subsection{Symmetry properties of the Green's functions}
\label{sec:symmetry}
Symmetries of the underlying Hamiltonian impose several exact constraints on the Green's functions, valid for any statistical phase $\theta$ and interaction strength $V$.
These relations anticipate several of the inversion properties discussed later in Sec.~\ref{sec:results}.

First, let us consider the effect of spatial inversion symmetry operator $\mc I$ and time-reversal symmetry operator $\mc T$ on the Green's functions.
The Hamiltonian is invariant under both symmetries for any statistical phase $\theta$ and interaction strength $V$.

The inversion symmetry operator $\mc I$ maps any lattice site to the mirror site with respect to the center of the chain, $j\to j' = L-j+1$, where primes denote the inverted site in the following.
Thus, an anyon operator at site $j$ is mapped under $\mc I$ to a similar anyon operator acting at the site $j'$, e.g.,~$\mc I a_j \mc I^\dag = a_{j'}$.
Inserting the identity $\id=\mc I^\dag\mc I$ inside the correlation functions determines that Green's functions at opposite statistical phases satisfy the relation
\begin{equation}\label{eq:Ginv}
G_{jk}(t;\theta,V) = G_{j'k'}(t;-\theta,V).
\end{equation}
The absence of a superscript on the Green's function indicates that the relation applies for any greater, lesser, or retarded Green's function.
Equation~\eqref{eq:Ginv} implies that the Green's functions are spatially inversion symmetric for bosons ($\theta=0$) and fermions ($\theta=\pi$) for any interaction strength $V$.

Time-reversal symmetry $\mc T$ acts as complex conjugation in the anyon Fock basis.
Inserting the identity $\id=\mc T^{-1} \mc T$ in the correlation functions yields the relation between Green's functions at opposite statistical phases when exchanging the spatial indices,
\begin{equation}
G_{jk}(t;\theta,V) = G_{kj}(t;-\theta,V).
\end{equation}
Combining the two properties results in Ref.~\cite{Wang2022}
\begin{equation}\label{eq:IT}
G_{jk}(t;\theta,V) = G_{k'j'}(t;\theta,V).
\end{equation}
By setting $k=j$, it follows that the equal-site Green's functions $G_{jj}(t;V,\theta)$ are inversion symmetric for any $\theta$ and $V$, which explains spatial inversion symmetry for the density of states in Fig.~\ref{fig:A_0_omega_Vs}.
The Green's functions also obey the conjugation relation, which relates Green's functions to their complex conjugates,
\begin{equation}\label{eq:conj}
G_{jk}(t;\theta,V) = -G_{kj}^*(-t;\theta,V).
\end{equation}

In order to understand the effect of interactions, we consider, similar to Ref.~\cite{Liu2018}, the effect of the odd number parity operator
\begin{equation}
\mc P = e^{i\pi \sum_{j} n_{2j+1}}, \quad \mc P a_j \mc P^\dag = (-1)^j a_j.
\end{equation}
The operator $\mc P$ commutes with the interaction term in the Hamiltonian, but anticommutes with the hopping term,
\begin{equation}
\mc P H(V) \mc P^\dag = -H(-V).
\end{equation}
Inserting identity operators as $\id=\mc P^\dag\mc P$ in the correlation functions yields the relation
\begin{equation}\label{eq:P}
G_{jk}(t;\theta,V) = (-1)^{j-k} G_{jk}(-t;\theta,-V).
\end{equation}

Using Eqs.~\eqref{eq:IT}, \eqref{eq:conj}, and \eqref{eq:P}, we obtain the relation between the Green's functions at opposite interaction strengths,
\begin{equation}\label{eq:GVs}
G_{jk}(t;\theta,V) = (-1)^{j-k+1} G_{j'k'}^*(t;\theta,-V).
\end{equation}
A particular consequence of the above relation is the inversion symmetry seen at $V=0$ in Fig.~\ref{fig:G_x_t_Vs}.
By setting $k=0$ as the reference site in the middle of the chain,
Eq.~\eqref{eq:GVs} indicates that the Green's functions are inversion symmetric up to a phase factor,
\begin{equation}\label{eq:V0_sym}
|G_{j0}(t;\theta,0)| = |G_{-j0}(t;\theta,0)|,
\end{equation}
i.e., $|G^>(x,t)| = |G^>(-x,t)|$ in Fig.~\ref{fig:G_x_t_Vs}.
At finite interactions $V$, the inversion symmetry is broken for any $\theta\neq 0,\pi$, as we will see in the following sections.
More details about the procedure to obtain the symmetry relations as well as additional numerical checks are provided in Appendix~\ref{app:symmetries}.

\section{Results}\label{sec:results}
\begin{figure}[t]
\centering
\includegraphics[width=\columnwidth]{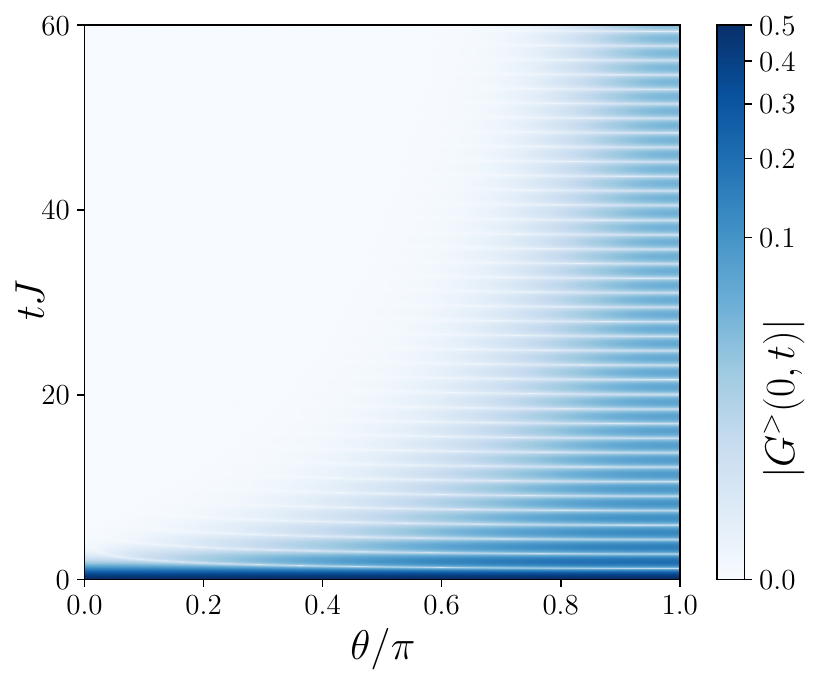}
\caption{The dependence of the free-anyon ($V=0$) Green's function $|G^>(0,t)|$ on time and the statistical phase $\theta$ from exact diagonalization simulations for $L=301$ sites. 
The data are power-law normalized to better visualize the small values.}
\label{fig:G_0_t_V_0}
\end{figure}
\subsection{Free-anyon Green's functions}
\label{sec:greens_functions}

\begin{figure*}[t]
\centering
\includegraphics[width=1.0\textwidth]{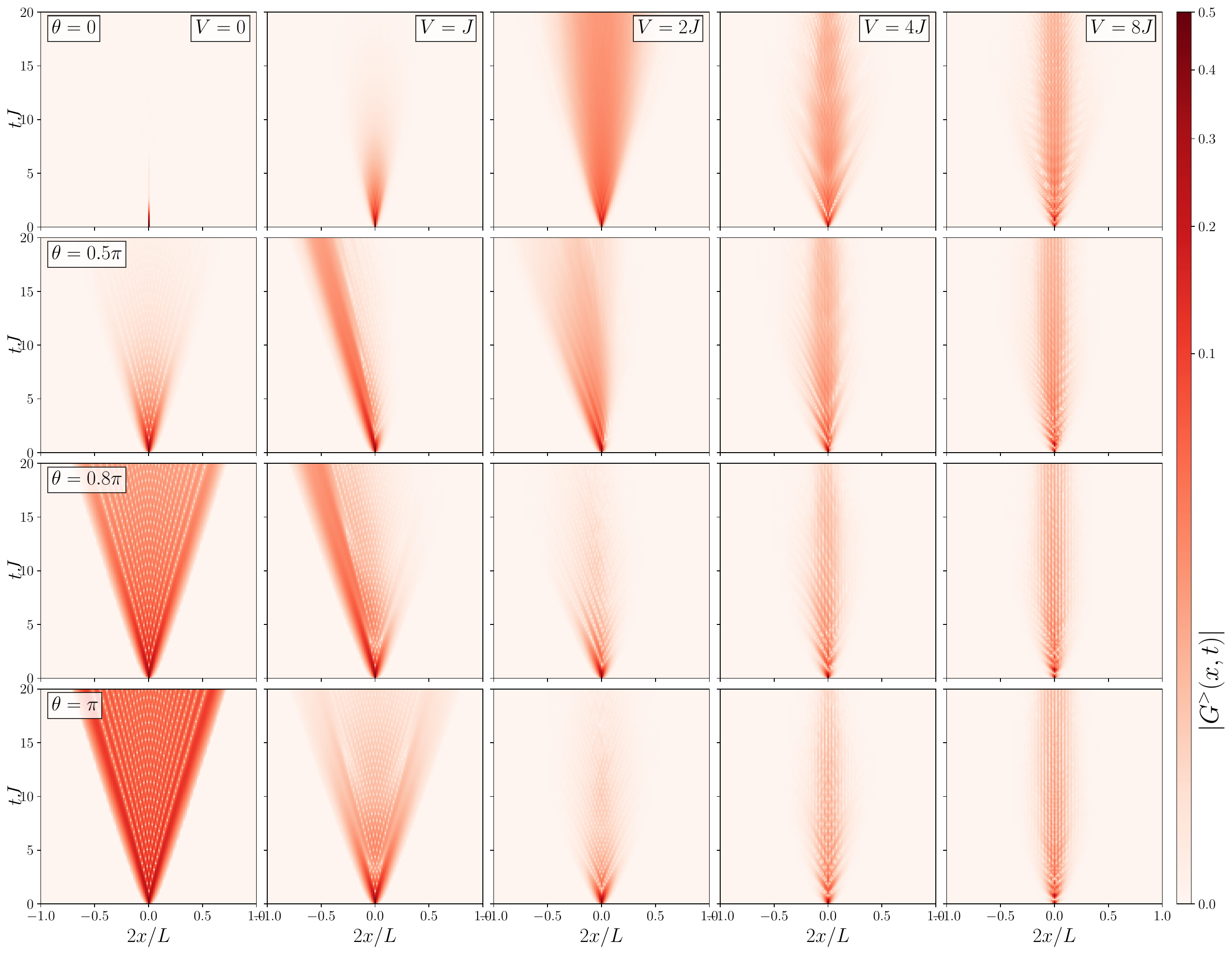}
\caption{The greater Green's function $|G^>(x,t)|$ evolution for five interaction strengths $V$ and four statistical phases.
From top to bottom row, the panels display $\theta=0.0$ (bosons), $\theta=0.5\pi$, $\theta=0.8\pi$, and $\theta=\pi$ (fermions), while columns are labeled by the interaction strength $V$.
The results are obtained for an $L=129$-site system using Fock-Liouville TEBD simulations with bond dimension $\chi=128$.}
\label{fig:G_x_t_Vs}
\end{figure*}

The natural starting point in the analysis is the noninteracting case $V=0$.
For the following discussion, the focus is on the greater Green's function introduced in Sec.~\ref{sec:green_definitions} by fixing the reference site at the center of the chain, $k=0$, and writing
\begin{equation}
G^{>}(x,t)= -i\langle a_x(t) a_0^\dag(0) \rangle,
\label{Eq:G_x_t}
\end{equation}
and the equal-time Green's function satisfies $G^{>}(x,0) = -\frac i2\delta_{x,0}$ for any $\theta$, which follows from the infinite-temperature limit.

In the limit cases $\theta=0$ (bosons) and $\theta=\pi$ (fermions), analytical results are available at infinite temperature, as we review in the following two subsections.
Surprisingly, for generic statistical phase $0<\theta<\pi$, there are no closed analytical solutions known in the noninteracting limit.
The problem is nevertheless amenable to several efficient and quasiexact numerical approaches at arbitrary temperatures.
From a Fredholm determinant representation of the Green's functions~\cite{Patu2015}, one can efficiently compute the dynamical correlations for large system sizes $L\sim\mc O(100)$.
Alternatively, a solution based on properties of quadratic fermionic Hamiltonians allows one to determine the dynamical correlations within machine precision for comparable system sizes~\cite{Wang2022}.
At finite interactions $V$ and time $t$, the correlations are computed using several methods~(see Appendix~\ref{app:methods}), benchmarked against exact solutions from exact diagonalization (ED) for small systems.

\subsubsection{Free fermions, \texorpdfstring{$\theta=\pi$}{theta=pi}}\label{sec:fermionic_limit}

In the free-fermion limit $\theta=\pi$, the quadratic Hamiltonian is diagonalized exactly in the infinite-chain limit in momentum space.
The Heisenberg equations of motion for the fermion operators are solved exactly, yielding
\begin{equation}\label{eq:ct_j}
c_j(t) = \sum_m i^{j-m} J_{m-j}(2Jt)c_m,
\end{equation}
where $J_n$ are Bessel functions of the first kind, and the sum runs over all sites $m$ of the chain.
The fermion operators $c_j$ without argument are evaluated at time $t=0$.
Therefore, the greater Green's function at $T=\infty$ follows as
\begin{align}\label{eq:G_fermion}
G^{>}_{jk}(t;\theta=\pi) &= -i \sum_{m} i^{j-m}J_{m-j}(2Jt)\avg{c_m c_k^\dag} \nonumber \\
&= \frac{i^{j-k-1}}{2}J_{k-j}(2Jt).
\end{align}
In particular, in the asymptotic long-time limit, the Green's function has periodic oscillations and decays as a power law in time,
\begin{equation}\label{eq:G_fermion_asympt}
G^{>}_{jk}(t;\theta=\pi) \sim \frac{i^{j-k-1}}{\sqrt{4\pi Jt}}
\cos\big(2Jt- \frac{k-j}{2}- \frac \pi 4\big).
\end{equation}

For the local correlator used in Figs.~\ref{fig:G_0_t_V_0} and~\ref{fig:G_0_t_limits}(a), Eq.~\eqref{eq:G_fermion} reduces to $G^{>}(0,t;\theta=\pi)=-\frac{i}{2}J_0(2Jt)$.
Figure~\ref{fig:G_0_t_V_0} shows that the ED data at $\theta=\pi$ follows this oscillatory Bessel-function form over the full simulated time window.
In real space, the fermionic $V=0$ panel in Fig.~\ref{fig:G_x_t_Vs} exhibits a ballistic and inversion-symmetric light cone, with fronts propagating at velocity $v=2J$.
The same exact fermionic result is highlighted in Fig.~\ref{fig:G_0_t_limits}(a), where the dashed curve displays the expected oscillatory decay with algebraic envelope $|G^>(0,t;\theta=\pi)|\simeq 1/\sqrt{4\pi Jt}$.

\subsubsection{Free bosons, \texorpdfstring{$\theta=0$}{theta=0}}\label{sec:bosonic_limit}
In the free bosonic limit $\theta=0$, the Green's function $G^{>}_{jk}(t;\theta=0)$ is strongly localized in time and space as
\begin{equation}\label{eq:G_boson}
G^{>}_{jk}(t;\theta=0) = -\frac{i}{2} e^{-J^2 t^2}\delta_{jk}.
\end{equation}
That is, no correlations are present between distinct sites.
This unexpected result was first conjectured~\cite{Sur1975} and then analytically proved~\cite{Brandt1976,Capel1977} in the 1970s [see also Ref.~\cite{Goehmann2020} for leading high-temperature corrections].
Figure~\ref{fig:G_0_t_V_0} illustrates that the ED data at $\theta=0$ follow the predicted Gaussian decay.
In real space, the bosonic $V=0$ panel in Fig.~\ref{fig:G_x_t_Vs} shows the absence of a ballistic light cone, its spreading being strongly suppressed.
The equal-site correlations decay rapidly in time according to the expected Gaussian form in Eq.~\eqref{eq:G_boson}, as emphasized in Fig.~\ref{fig:G_0_t_limits}(a) for the data set with $\theta=0$.

\subsubsection{Free anyons, \texorpdfstring{$0<\theta<\pi$}{0<theta<pi}}\label{sec:generic_phase}

\begin{figure}[t]
\centering
\includegraphics[width=\columnwidth]{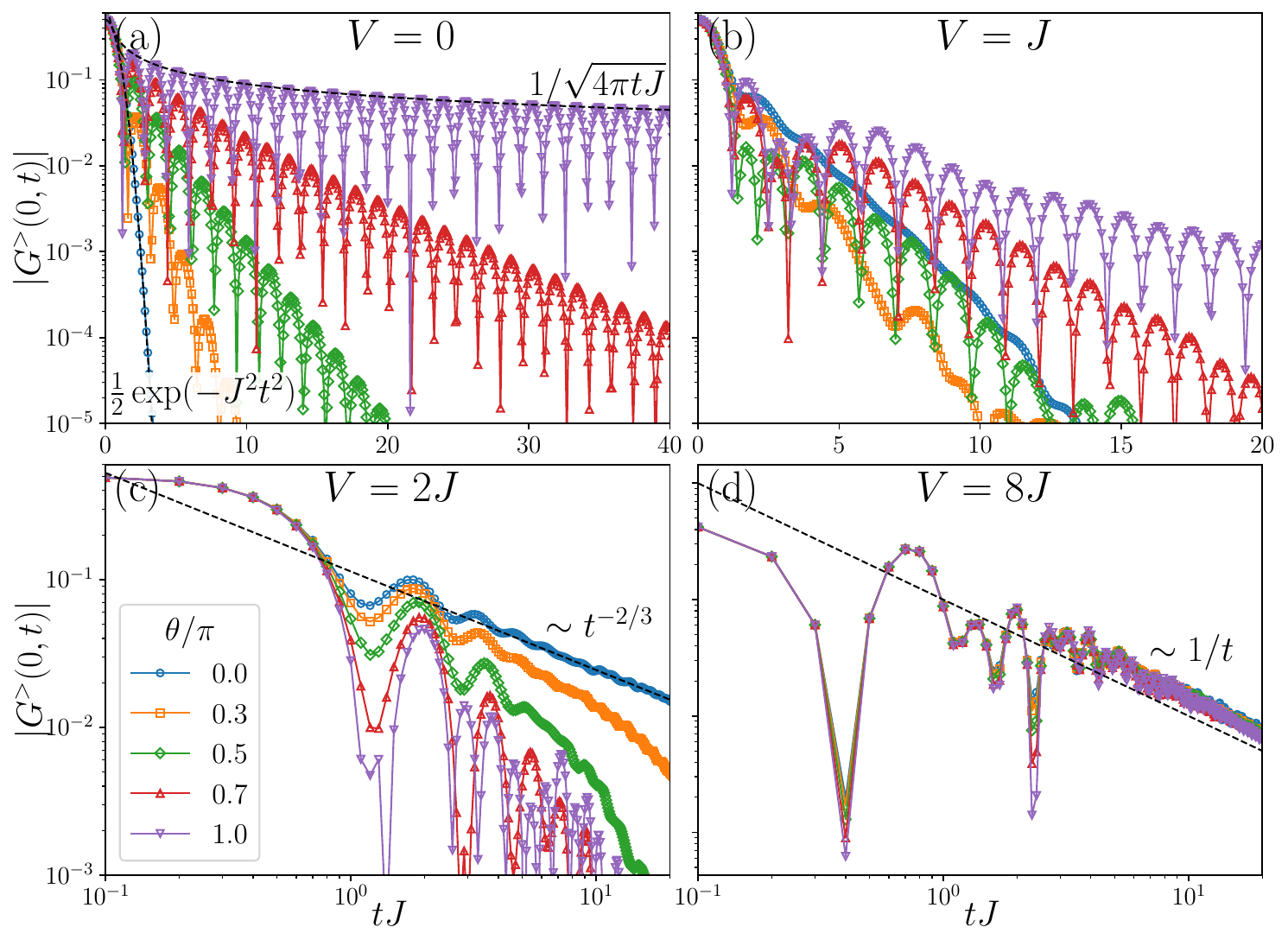}
\caption{Time dependence of the local greater Green's function amplitude $|G^>(0,t)|$ for several statistical phases $\theta$ at (a) $V=0$, (b) $V=J$, (c) $V=2J$, and (d) $V=8J$.
The labeled black dashed lines in panel (a) mark the analytical results: Gaussian decay for bosons~\eqref{eq:G_boson} and the power-law decay for fermions~\eqref{eq:G_fermion_asympt}.
The dashed line in panel (c) indicates the KPZ power-law decay $|G^>(0,t;\theta=0)|\sim t^{-1/z}$, with $z=3/2$ at $V=2J$.
The dashed line in panel (d) indicates the strong-coupling asymptotic decay $|G^>(0,t)|\sim t^{-1}$, valid for any $\theta$.
Panels (a) and (b) are plotted on a semilog timescale to better visualize the exponential decay, while panels (c) and (d) are plotted on a log-log scale.
The legend in panel (c) is shared by all panels.
}
\label{fig:G_0_t_limits}
\end{figure}

We now turn to intermediate statistical phases $0<\theta<\pi$, which interpolate between the two exactly solvable endpoints discussed above.
In this regime, no closed analytical expression is available even at $V=0$, and we therefore rely on exact diagonalization~\cite{Wang2022} and TEBD simulations~\cite{Vidal2003} performed using the \texttt{ITensors} library~\cite{Fishman2022}.
The numerical results connect smoothly the bosonic and fermionic limits and recover the analytical expressions~\eqref{eq:G_boson} and~\eqref{eq:G_fermion} at the boundaries in Figs.~\ref{fig:G_0_t_V_0},~\ref{fig:G_x_t_Vs}, and~\ref{fig:G_0_t_limits}(a).

The results in Fig.~\ref{fig:G_x_t_Vs} show that for almost any angle $\theta$ there is ballistic spreading of correlations, but the spatial extent becomes strongly attenuated as the bosonic limit is approached.
The velocity of the correlation front is set by the hopping scale, $v=2J$, similarly to the free-fermion limit.
In the bosonic limit, in contrast, destructive interference leads to the exponential suppression of correlations in time encoded in Eq.~\eqref{eq:G_boson}.

Figure~\ref{fig:G_0_t_limits}(a) shows how the decay of the local Green's function $|G^>(0,t)|$ interpolates between the Gaussian bosonic behavior at $\theta=0$ and the fermionic power-law decay at $\theta=\pi$.
For most intermediate statistical phases $\theta \gtrsim 0.1$, the decay is approximately described by an exponential envelope 
\begin{equation}
    \label{eq:G0t_decay}
|G^>(0,t)| \sim e^{-\alpha(\theta) tJ},
\end{equation}
with an exponent that depends on the statistical phase $\theta$, together with pronounced time oscillations (see Table~\ref{tab:V0}).
Below $\theta\lesssim 0.1$, the decay is faster than exponential, as the system approaches the Gaussian decay of the bosonic local Green's function~\eqref{eq:G_boson}.
Close to the fermionic limit, $\alpha(\theta)\to 0$, the power-law decay $1/\sqrt{tJ}$  is recovered, as expected.
The oscillation frequency, however, remains essentially the same as in the fermionic limit; i.e., it is set by the hopping scale $J$, $\omega\approx 2J$.
This is apparent in Fig.~\ref{fig:G_0_t_V_0}, where the extrema and zero crossings for intermediate statistical phases occur on nearly the same time scale as for $\theta=\pi$.
Thus, varying $\theta$ primarily changes the decay envelope, while the characteristic oscillation scale continues to be set by the hopping dynamics.

\begin{table}[h!]
    \caption{
        Decay rate $\alpha(\theta)$ of $|G^>(0,t)|$~\eqref{eq:G0t_decay} for the data from Fig.~\ref{fig:G_0_t_limits}(a)  using exact diagonalization~(Appendix~\ref{sec:ED}).
    }
    \begin{tabular}{cc}
        \hline\hline
        $\theta/\pi$ & $\alpha(\theta)$ \\
        \hline
        $0.3$ & $1.0781\pm 0.0006$ \\
        $0.5$ & $0.510\pm 0.001$ \\
        $0.7$ & $0.204\pm 0.002$ \\
        \hline\hline
    \end{tabular}
    \label{tab:V0}
\end{table}

Crucially, in the infinite-temperature limit, the Green's functions possess inversion symmetry at $V=0$ [see Eq.~\eqref{eq:V0_sym}; Sec.~\ref{sec:symmetry}] for all statistical phases $\theta$.
This is in contrast to finite-temperature Green's functions~\cite{Patu2015,Wang2022}, where spatial inversion symmetry is broken for anyons with $0<\theta<\pi$ as a consequence of the nontrivial interplay between the thermal density matrix and the Jordan-Wigner strings.

\begin{figure*}[t]
\centering
\includegraphics[width=\textwidth]{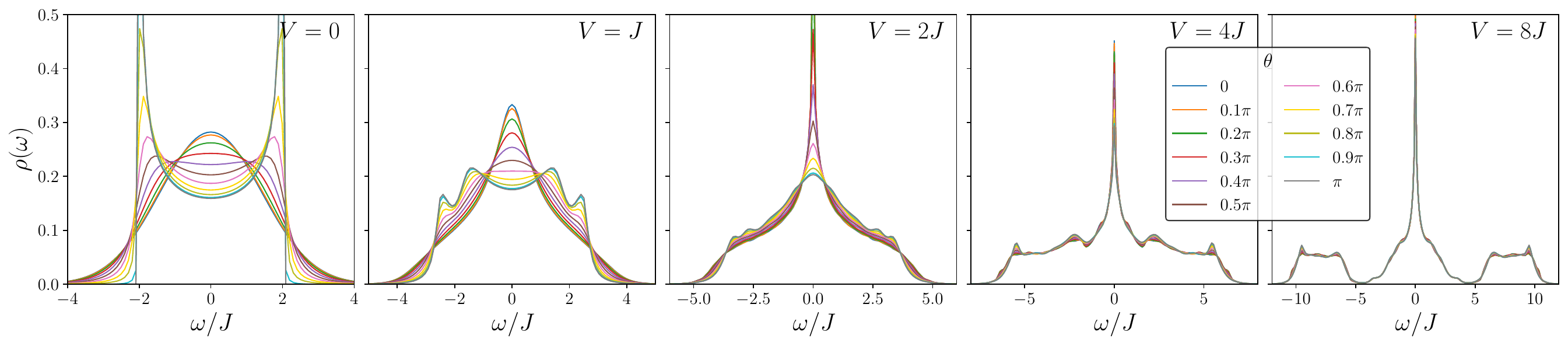}
\caption{Anyon density of states $\rho(\omega)$ in the presence of interactions for several statistical phases $\theta$ and interaction strengths $V$ from Fock-Liouville time-evolving block decimation (TEBD) simulations for $L=129$ sites and bond dimension $\chi=128$. The legend is shared by all panels.
}
\label{fig:A_0_omega_Vs}
\end{figure*}

\subsection{Interacting Green's functions}\label{sec:interacting_greens_functions}
\subsubsection{Two-site anyon problem}\label{sec:two_site_problem}

Let us now consider the effect of finite nearest-neighbor interactions $V\neq 0$ on the dynamical correlations.
Although the additional interaction term does not depend explicitly on the statistical phase $\theta$, it does not commute with the anyon operators.
This leads to a nontrivial effect: the breaking of spatial inversion symmetry in the correlation functions, as we show below.
The breaking of inversion symmetry is already apparent from a two-site anyon problem.
In this case, the Hamiltonian reads
\begin{equation}
\label{eq:2site_ham}
H = J(a_1^\dag a_2 + a_2^\dag a_1) + V \left(n_1-\frac12\right)\left(n_2-\frac12\right).
\end{equation}
The greater Green's function is computed exactly, and the on-site correlations are identical on both sites, $G^>_{jj}(t)=-\frac i2\cos(Jt)\cos(Vt/2)$.
The off-diagonal correlations read
\begin{align}
G^>_{12}(t) &= -\frac i2e^{-i\frac\theta 2}\sin(Jt)
\sin\left(\frac{Vt+\theta}{2}\right),\notag\\
G^>_{21}(t) &= -\frac i2 e^{+i\frac\theta 2}\sin(Jt)\sin\left(\frac{Vt-\theta}{2}\right),
\end{align}
such that in general $G^>_{12}(t) \neq G^>_{21}(t)$.
For special cases, the inversion symmetry is restored.
This happens at $V=0$ for any $\theta$, as discussed above.
It is also restored for bosons ($\theta=0$) and fermions ($\theta=\pi$) at any $V$.
The two-site solution identifies the microscopic mechanism responsible for inversion-symmetry breaking. Since the many-body Hamiltonian is a sum of identical nearest-neighbor terms, this local mechanism naturally extends to larger lattices and gives rise to the inversion asymmetry observed numerically.

To move beyond the exactly solvable two-site problem, and to investigate the inversion symmetry breaking in larger systems in more detail, we study finite open chains using TEBD.
The spatiotemporal profiles in Fig.~\ref{fig:G_x_t_Vs} show the main qualitative effect of interactions at $T=\infty$. 
For anyons with $0<\theta<\pi$, finite nearest-neighbor interactions break the inversion symmetry of the Green's functions, whereas bosons and fermions remain inversion symmetric for all $V$.
The asymmetry is most pronounced at intermediate couplings $V\sim J$, where hopping and interactions compete most effectively.
For larger $V$, the dynamics becomes more localized, the dependence on the statistical phase weakens, and the profiles in Fig.~\ref{fig:G_x_t_Vs} approach an approximately inversion-symmetric form.
The local decay $|G^>(0,t)|$ shown in Fig.~\ref{fig:G_0_t_limits} provides a complementary view, as it shows how the decay evolves with $V$ for different $\theta$.
To extract the decay rates, we fit the data for $|G(0,t)|>10^{-6}$, which is the threshold for the numerical noise in the TEBD simulations for our simulation parameters.
The initial transient dynamics at $tJ\lesssim4$ is excluded from the fits.

\subsubsection{Interacting bosons, \texorpdfstring{$\theta=0$}{theta=0}}\label{sec:interacting_bosons}
In the bosonic limit $\theta=0$, the local Green's functions display a sharp crossover at $V=2J$ between different decay laws.
For $0<V<2J$, one finds that
\begin{equation}
|G^>(0,t;\theta=0)| \sim e^{-\alpha_0(V)t}, \quad 0<V<2J,
\end{equation}
with $\alpha_0(V)>0$ decreasing as $V\to 2J$ (see Table~\ref{tab:inter}).
Due to the exponential decay, the Green's function behavior is extracted up to a time $20/J$ in Figs.~\ref{fig:G_0_t_limits}(b) and~\ref{fig:G_0_t_limits}(c).
At the crossover point $V=2J$, one finds the superdiffusive decay
\begin{equation}
|G^>(0,t;\theta=0)| \sim t^{-1/z}, \quad z=3/2, \quad V=2J,
\end{equation}
where the fitted exponent has a standard error of approximately $10^{-3}$.
This behavior can be understood in terms of the mapping to the XXZ chain, where the point $V=2J$ corresponds to the SU(2)-symmetric Heisenberg point, and the transverse spin correlations match the longitudinal spin correlations, and  the system is known to display superdiffusive spin transport with dynamical exponent $z=3/2$~\cite{Ljubotina2017, Ljubotina2019}.

For $V>2J$, in contrast, the local Green's function remains algebraic and displays a power-law decay with an exponent that approaches $z=1$ at large $V$,
\begin{equation}
|G^>(0,t;\theta=0)| \sim t^{-1}, \quad V>2J.
\end{equation}
For example, at $V=8J$, the data in Fig.~\ref{fig:G_0_t_limits} yield a dynamical exponent $z=1.023\pm 0.022$ when data are fitted over the time window $tJ\in[4,15]$.

\subsubsection{Interacting fermions, \texorpdfstring{$\theta=\pi$}{theta=pi}}\label{sec:interacting_fermions}

In the fermionic limit, the free result discussed in Sec.~\ref{sec:fermionic_limit} gives the algebraic decay
\begin{equation}
|G^>(0,t;\theta=\pi)| \sim t^{-1/2}, \quad V=0,
\end{equation}
but Figs.~\ref{fig:G_0_t_limits}(b) and~\ref{fig:G_0_t_limits}(c) show that this behavior is rapidly replaced by an approximately exponential decay once a finite interaction is switched on.
For weak but finite interactions, the numerical data are well described over the accessible time window by the exponential envelope
\begin{equation}
|G^>(0,t;\theta=\pi)| \sim e^{-\alpha_\pi(V)tJ}, \quad 0<V\lesssim 2J,
\end{equation}
with an effective decay rate $\alpha_\pi(V)>0$ that vanishes as $V\to 0$ (see Table~\ref{tab:inter}).
Unlike in the bosonic case, the fermionic data do not display a sharp feature at $V=2J$; rather, the decay evolves smoothly with $V$.
This is due to the fermionic Green's function corresponding to a nonlocal string-dressed correlator in the spin language that is not constrained by SU(2) symmetry in the same direct manner.
Moreover, the smooth decay is consistent with the singular character of the free-fermion point for one-body observables: Away from $V=0$, interactions introduce many-body dephasing that damps the coherent Bessel-function oscillations characteristic of the noninteracting limit, qualitatively in line with the interaction-induced relaxation of one-body observables discussed in Ref.~\cite{Wright2014}.
For larger $V$, the decay crosses over again toward the same strong-coupling asymptotic law found for generic anyons,
\begin{equation}
|G^>(0,t;\theta=\pi)| \sim t^{-1}, \quad V\gg J,
\end{equation}
as seen in Fig.~\ref{fig:G_0_t_limits}(d).
We interpret this regime as the approach to the atomic limit, where the low-frequency part of the local spectrum develops the logarithmic 
singularity discussed below for the density of states, whose Fourier transform yields the long-time $1/t$ tail.

\begin{table}
\caption{
    The decay rates $\alpha(\theta,V)$ of $|G^>(0,t)|$ shown in Figs.~\ref{fig:G_0_t_limits}(b) and~\ref{fig:G_0_t_limits}(c) when it is fitted to $\exp(-\alpha(\theta,V)tJ)$ for $V=J$ and $V=2J$ from TEBD simulations.
}
\begin{tabular}{ccc|ccc}
\hline\hline
$V$ & $\theta/\pi$ & $\alpha$ & $V$ & $\theta/\pi$ & $\alpha$ \\
\hline
\multirow{5}{1.2em}{$J$} & $0.0$ & $0.701\pm 0.009$ & \multirow{5}{1.2em}{$2J$} & $0.0$ & ---\\
& $0.3$ & $0.841\pm 0.007$ & & $0.3$ & $0.117\pm 0.001$ \\
& $0.5$ & $0.573\pm 0.009$ & & $0.5$ & $0.277\pm 0.003$ \\
& $0.7$ & $0.446\pm 0.003$ & & $0.7$ & $0.52\pm 0.03$ \\
& $1.0$ & $0.27\pm 0.02$ & & $1.0$ & $0.63\pm 0.06$ \\ 
\hline\hline
\end{tabular}
\label{tab:inter}
\end{table}

\subsubsection{Interacting anyons, \texorpdfstring{$0<\theta<\pi$}{0<theta<pi}}\label{sec:interacting_anyons}
For intermediate statistical phases $0<\theta<\pi$, the free-limit behavior already consists of oscillations dressed by an approximately exponential envelope, as discussed in Sec.~\ref{sec:generic_phase}.
Figures~\ref{fig:G_0_t_limits}(b) and~\ref{fig:G_0_t_limits}(c) show that this structure survives at weak interactions, where the local Green's function is still well described over the accessible time window by
\begin{equation}
|G^>(0,t;\theta)| \sim e^{-\alpha(V,\theta)tJ}, \quad 0<V\lesssim J, \quad 0<\theta<\pi,
\end{equation}
with an effective decay rate $\alpha(V,\theta)>0$ that connects smoothly to the free-limit decay rate as $V\to 0$.
Extracted decay rates for the data shown in Figs.~\ref{fig:G_0_t_limits}(b) and~\ref{fig:G_0_t_limits}(c) are summarized in Table~\ref{tab:inter}.
Thus, unlike in the fermionic case, finite interactions do not introduce a new functional form at small $V$, but rather renormalize the decay rate and progressively damp the oscillations.
This sharp separation between weak, intermediate, and strong-coupling behaviors is replaced by a smoother crossover.
The point $V=2J$ no longer singles out a distinct critical behavior, and the oscillatory decay evolves continuously from the free-limit behavior discussed above toward the universal strong-coupling asymptotic
\begin{equation}\label{eq:strong_V_decay}
|G^>(0,t;\theta)| \sim t^{-1}, \quad \forall \theta,
\end{equation}
as shown in Fig.~\ref{fig:G_0_t_limits}(d).
As in the fermionic case, we interpret this large-$V$ regime as the approach to the atomic limit, where the low-frequency logarithmic singularity of the local density of states generates the long-time $1/t$ tail.

\subsection{Density of states}\label{sec:density_of_states}

\begin{figure*}[t]
\centering
\includegraphics[width=\textwidth]{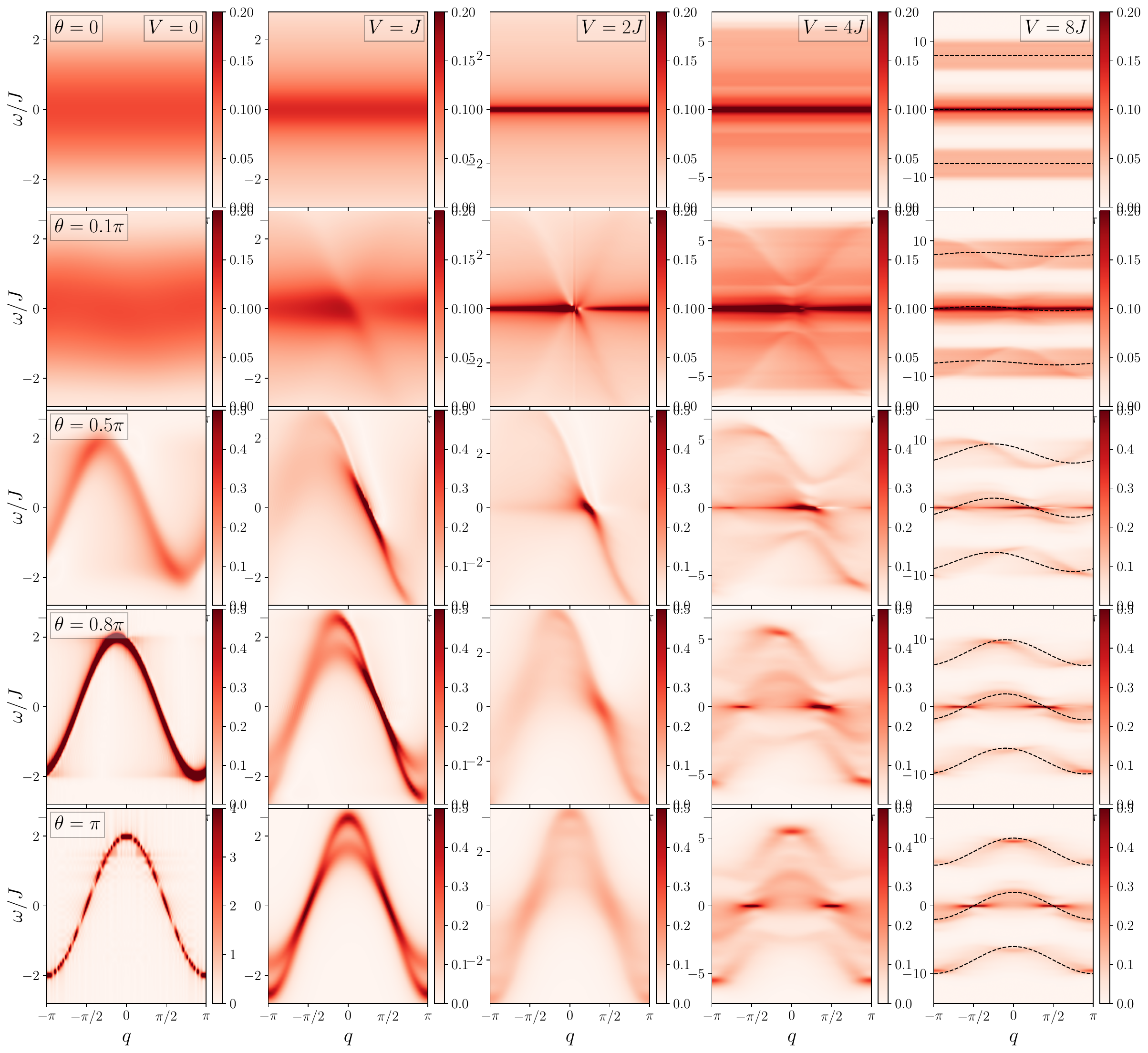}
\caption{Spectral function $A(q,\omega)$ for several statistical phases using TEBD in Fock-Liouville space for a system of size $L=129$ sites and bond dimension $\chi=128$.
The rows are labeled by the statistical phase $\theta$, while columns are labeled by the interaction strength $V$.
The dashed lines in the panels corresponding to $V=8J$ indicate the mean-field dispersion from Eq.~\eqref{eq:omega_m}}
\label{fig:A_q_omega}
\end{figure*}
The local density of states $\rho(\omega)$, shown in Fig.~\ref{fig:A_0_omega_Vs}, is computed from Eq.~\eqref{eq:dos}.
We first discuss the free limit $V=0$, where the line shape evolves smoothly from the bosonic to the fermionic case as the statistical phase is increased.
For bosons, one obtains the Gaussian form
\begin{equation}
\label{eq:dos_boson}
\rho(\omega;\theta=0) = e^{-\omega^2/4J^2}/\sqrt{4\pi J^2},
\end{equation}
and this broad single-peak structure persists for small $\theta$.
As $\theta$ increases, spectral weight is transferred toward the band edges $\omega=\pm 2J$, side peaks develop, and in the fermionic limit one recovers the familiar Van Hove singularities,
\begin{equation}
\rho(\omega;\theta=\pi) = \frac{1}{\pi\sqrt{4J^2-\omega^2}}.
\end{equation}

Finite interactions redistribute the same spectral weight while preserving the sum rule $\int d\omega\,\rho(\omega)=1$.
The resulting structure is easiest to understand in the atomic limit $J/V\to 0$.
At $T=\infty$, each site is empty or occupied with probability $1/2$, so the local energy cost of adding or removing a particle is determined only by the occupations of the two neighboring sites.
For an initially empty site $j$, adding a particle changes the interaction energy by
\begin{equation}
\Delta E=V(n_{j+1}+n_{j-1}-1).
\end{equation}
Thus, $\Delta E$ can take the three values $-V$, $0$, and $+V$, corresponding to the neighboring configurations $(0,\,0)$, $(0,\,1)$ or $(1,\,0)$, and $(1,\,1)$.
Their probabilities are $1/4$, $1/2$, and $1/4$, which directly explains the $1{:}2{:}1$ weight ratio of the three structures centered near $\omega\approx -V$, $0$, and $+V$ in Fig.~\ref{fig:A_0_omega_Vs}.
The same set of energies and weights is obtained for particle removal, so the local density of states keeps the same three-peak pattern.
Finite hopping introduces a broadening of the peaks, leading to the formation of bands of width $\simeq 4J$ around the atomic-limit energies.
Thus, at $V=8J$ the numerical spectra show the well-separated three-band structure, separated by gaps of size $\simeq 4J$.
We have checked that the integrated density of states within each band at $V=8J$ follows the $1{:}2{:}1$ ratio with good accuracy.

At strong interactions, the Green's function is independent of the statistical phase $\theta$ in the long-time limit according to Eq.~\eqref{eq:strong_V_decay}.
This produces a logarithmic divergence of $\rho(\omega)$ at $\omega=0$ for all $\theta$, as visible in Fig.~\ref{fig:A_0_omega_Vs}.
In the same regime, the dependence on the statistical phase becomes much weaker, again consistent with the atomic-limit picture $V\to\infty$, where the statistical phase drops out and the local spectrum reduces to three peaks at $\omega=0$ and $\omega=\pm V$.

\begin{figure*}[t]
\centering
\includegraphics[width=0.95\textwidth]{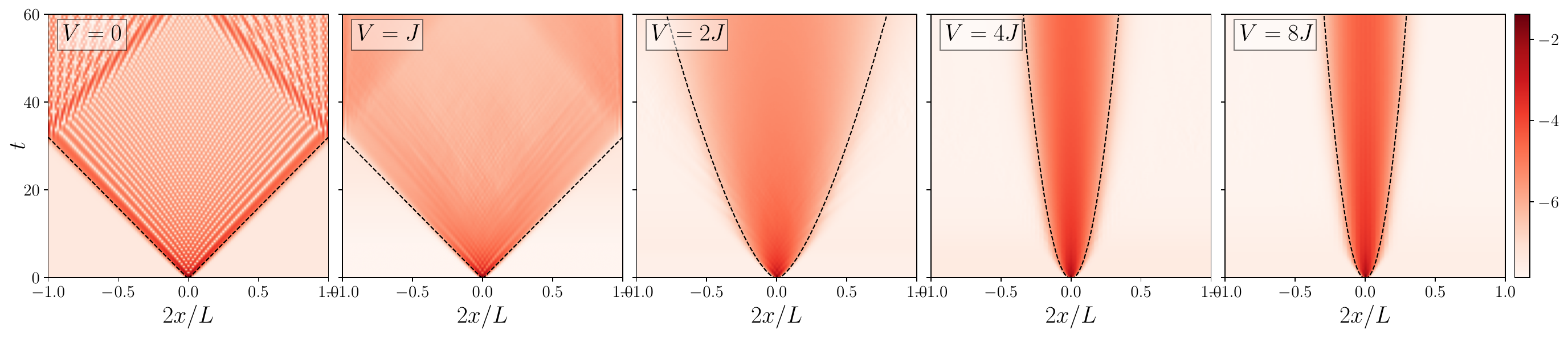}
\caption{The connected density-density correlation function $\ln[C^c_{j0}(t)]$ at infinite temperature is independent of the statistical phase $\theta$.
The dashed black lines represent the scaling of the correlation front $\Delta x\sim t^{1/z}$, with $z=1$ for ballistic transport ($V=0$ and $J$), $z=3/2$ for superdiffusive transport ($V=2J$), and $z=2$ for diffusive transport ($V=4J$ and $8J$).
Reflection from the system boundaries is visible at $V=0$ and $V=J$.
The data were obtained from operator-evolution TEBD for $L=129$ sites and bond dimension $\chi=128$. 
}
\label{fig:NN_anyon}
\end{figure*}

\subsection{Momentum-resolved spectral function}\label{sec:spectral_function}

The momentum-resolved spectral function $A(q,\omega)$ shown in Fig.~\ref{fig:A_q_omega} is particularly transparent in the two exactly solvable free limits.
For bosons, the spectral function at $V=0$ is momentum independent and reduces to the same Gaussian line shape as the local density of states~\eqref{eq:dos_boson},
\begin{equation}
A(q,\omega;\theta=0,V=0)
= e^{-\omega^2/4J^2}/\sqrt{4\pi J^2},
\end{equation}
such that every momentum carries the same broad incoherent spectral weight.
For free fermions, in contrast, the spectral function collapses onto the single-particle dispersion,
\begin{equation}\label{eq:A_q_omega_fermion}
A(q,\omega;\theta=\pi,V=0) = \delta(\omega-\omega_q),
\quad \omega_q = 2J\cos q,
\end{equation}
which describes infinitely long-lived quasiparticle modes with well-defined energy.

For intermediate statistical phases $0<\theta<\pi$ at $V=0$, the spectral function interpolates continuously between these two limits.
The broad bosonic Gaussian evolves into a dispersive structure whose maximum shifts with momentum, and already at $\theta=0.5\pi$ the result resembles the fermionic band, but displaced in momentum.
Since the energy spectrum itself is independent of $\theta$, 
the asymmetric redistribution of the spectral weight at $q$ and $-q$ is due entirely to the matrix elements of the string-dressed anyon operators.

In order to understand the structure of the spectral function, it is useful to write down the equations of motion for the retarded Green's function $G^R_{j0}(\omega)$ and seek a mean-field solution.
Using Zubarev's method and notations~\cite{Zubarev1960}, the equation reads 
\begin{equation}
(\omega+i0^+)G_{j0}^R(\omega) = \zavg{[a_j,H];a_0^\dag} + \avg{[a_j,a_0^\dag]_\theta},
\end{equation}
with the anyon commutator $[\cdot,\cdot]_\theta$ from Eq.~\eqref{eq:Gtjk} and
\begin{equation}
G_{j0}^R(\omega) \equiv \zavg{a_j;a_0^\dag} 
= -i\int_0^\infty dt e^{i\omega t}\avg{[a_j(t),a_0^\dag(0)]_\theta}.
\end{equation}
After evaluating the commutator between the Hamiltonian and the anyon operators, one obtains
\begin{widetext}
\begin{equation}\label{eq:eom}
(\omega+i0^+)G_{j0}^R(\omega) =
\zavg{a_{j+1}e^{i(\pi-\theta)n_j};a_0^\dag} 
+\zavg{a_{j-1}e^{i(\theta-\pi)n_j};a_0^\dag}
+V(\zavg{a_jn_{j+1};a_0^\dag} + \zavg{a_j n_{j-1};a_0^\dag} - G_{j0}^R(\omega))
+\delta_{j0}.
\end{equation}
\end{widetext}
The equations are not closed due to the generation of higher-order correlators both in hopping and interaction terms in the form of couplings between density operators and anyon operators.
Nevertheless, as interactions enter the strong-coupling regime $V\gg J$, the three-band structure seen most clearly in the last column of Fig.~\ref{fig:A_q_omega} can be understood from a simple mean-field expansion around the atomic limit.
The energy cost of adding or removing a particle at site $j$ in a frozen local background is controlled by the effective field
\begin{equation}
h_j = V(n_{j-1}+n_{j+1}-1) = \nu V,\quad \nu=0,\pm1.
\end{equation}
Additionally, the exponentials in Eq.~\eqref{eq:eom} are expanded and the local density operator is decoupled by replacing $n_j$ with its average value $\avg{n_j}=1/2$ at $T=\infty$.
The fermion limit is particularly straightforward as the hopping terms do not generate higher-order correlators and the approximation $n_j\mapsto \avg{n_j}$ is not required.

At infinite temperature, the three local environments occur with probabilities
$p_{-1}=p_{+1}=1/4$ and $p_0=1/2$.
Replacing the neighboring occupations by one of these static mean fields and Fourier transforming~Eq.~\eqref{eq:eom} gives the approximate retarded propagator
\begin{equation}
G^R_{\rm MF}(q,\omega) \simeq \sum_{\nu}
\frac{p_\nu}{\omega-\nu V+2J\sin(q-\frac\theta 2)\sin(\frac\theta 2)+i0^+},
\end{equation}
and therefore
\begin{equation}
A_{\rm MF}(q,\omega) \simeq \sum_{\nu} p_\nu\,
\delta\!\left[\omega-\nu V+2J\sin(q-\frac\theta 2)\sin(\frac\theta 2)\right].
\label{eq:Aqomega_MF}
\end{equation}
Thus, in the large-$V$ expansion,
\begin{equation}
\omega_\nu(q)\simeq \nu V-2J\sin(q-\frac\theta 2)\sin(\frac\theta 2),\quad \nu=0,\pm1,
\label{eq:omega_m}
\end{equation}
so the three atomic lines at $\omega=0$ and $\omega=\pm V$ broaden into bands.
Equation~\eqref{eq:Aqomega_MF} explains the main features of the last column of Fig.~\ref{fig:A_q_omega}. 
The central band originates from local environments with one occupied and one empty neighbor, while the upper and lower bands come from the $(1,\,1)$ and $(0,\,0)$ configurations, respectively.

For fermions, the absence of anyonic strings means that the mean field is exact at $V=0$, and one recovers as expected $A(q,\omega)=\delta(\omega-2J\cos q)$ from Eq.~\eqref{eq:A_q_omega_fermion}.
The strong features at $q=0$ and $q=\pm\pi$ are the ordinary one-dimensional band-edge singularities of the free-fermion branch, where the group velocity vanishes.
At strong interactions, the absence of anyonic strings in the hopping terms means that these do not mix local sectors, and the corrections that mix different local sectors are only of order $J^2/V$.
Thus, the mean-field approximation is also quite accurate at large $V$, where Eq.~\eqref{eq:omega_m} predicts the three-band structure with dispersions $\omega_\nu(q)\simeq \nu V+2J\cos q$, $\nu=0,\pm1$.
The bright spots at $q=\pm\pi$ in the upper and lower side bands are therefore again zone-boundary band-edge features of the shifted fermionic dispersions $\omega_{\pm}(q)$, enhanced by the same one-dimensional Van Hove mechanism.
In contrast, the strong signal in the central band near $q=\pm\pi/2$ is not a band-edge effect.
For the central fermionic branch, $\omega_0(q)=2J\cos q+\mc O(J^2/V)$ crosses zero at $q=\pm\pi/2$, so these points mark the intersection of that branch with the strong low-frequency spectral weight developing around $\omega=0$ in the large-$V$ regime.
The pronounced intensity there is thus better interpreted as a zero-energy crossing feature of the central fermionic branch, rather than as a Van Hove singularity.

As the statistical phase is varied from the fermion limit, the mixing between local sectors increases and the approximation $n_j\mapsto 1/2$ fails to capture the full structure of the spectral function.
The MF equations of motion predict a simple shift of the dispersion by $q\mapsto q-\theta/2$, which remains mostly correct for the central band $\omega_0$ (or for all bands at $\theta\simeq\pi$), but fails to capture the more complex structure of the sidebands at intermediate $\theta$.
Spectral-weight accumulation occurs, as in the fermionic case, when $\omega_0=0$ at $q\simeq \theta/2$ and $q\simeq \theta/2-\pi$.
The mean-field approximation~\eqref{eq:omega_m} predicts simple replicated bands, while the numerics show that for generic anyons the sidebands are strongly distorted and have energy-dependent momentum shifts.
Finally, for bosons, the mean-field equations yield three flat bands at $\omega=-V$, $0$, and $+V$, thus correctly capturing the momentum independence of the spectral functions.
Nevertheless, the mean-field treatment loses all information about the correct bandwidths for bosons.
In contrast, the numerics show that the three bands have approximately the same width $4J$, independent of $\theta$.

\subsection{Density-density correlations}
\label{sec:density_density_correlations}
The density-density correlators convey a simpler message than the Green's functions.
Because the number operators $n_j$ do not contain Jordan-Wigner strings, $C_{jk}(t)$ and $C^c_{jk}(t)$ are completely independent of the statistical phase $\theta$.
Therefore, they isolate the dynamics of the conserved density.
In the bosonic language, they coincide with the longitudinal spin correlators of the XXZ chain, so at $T=\infty$ they are expected to display the standard XXZ transport regimes~\cite{Znidaric2011,Ljubotina2017,Ljubotina2019}: ballistic for $V<2J$, superdiffusive at the isotropic point $V=2J$, and diffusive for $V>2J$.
These results are not the primary contribution of the present work but instead serve as a benchmark demonstrating that, in the bosonic limit $\theta\to0$, our TEBD simulations correctly reproduce the established infinite-temperature XXZ hydrodynamics.

\begin{figure}[t]
\centering
\includegraphics[width=0.9\columnwidth]{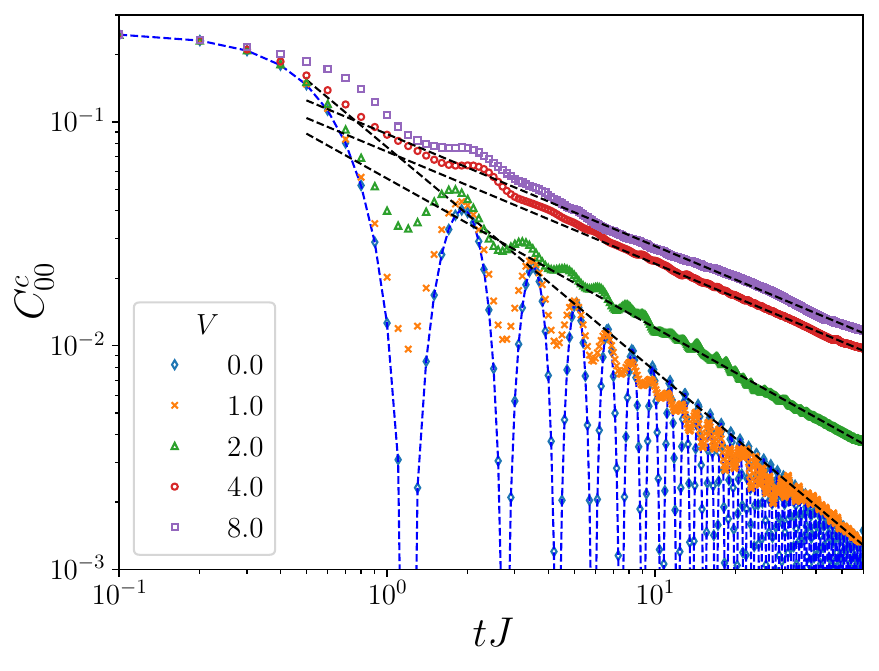}
\caption{Decay of the connected density-density correlator $C^c_{00}(t)$ at infinite temperature.
The numerical results, denoted by symbols, are slices at $x=0$ from the data shown in Fig.~\ref{fig:NN_anyon}.
The dashed blue line is the analytical result~\eqref{eq:nn_V0} for the free-anyon case $V=0$, interpolating perfectly the numerical data.
The long-time decay of the on-site correlations for $V>0$ is well fitted by (black dashed lines) $c(tJ)^{-1/z}$, with $c$ a fitting constant and ballistic $z=1$ for $V=J$, KPZ $z=3/2$ for the critical $V=2J$, and diffusive $z=2$ for $V=4J$ and $8J$. 
}
\label{fig:dens_dens_fit}
\end{figure}

The free limit $V=0$ can be treated analytically using Wick's theorem.
For the connected correlator, Eq.~\eqref{eq:ct_j} yields the Bessel-function representation
\begin{align}\label{eq:nn_V0}
C^c_{jk}(t) &= C_{jk}(t) - \avg{n_j(t)} \avg{n_k(0)} \notag\\
&= \frac14 J_{j-k}(2Jt)^2.
\end{align}
In the following, we fix the reference site at the center of the chain, $k=0$.
Equation~\eqref{eq:nn_V0} immediately shows that the density sector is identical for all statistical phases and that the correlations spread ballistically in the free case.
The analytical expression agrees with the numerical data in Figs.~\ref{fig:NN_anyon} and~\ref{fig:dens_dens_fit}, confirming the ballistic $z=1$ regime at $V=0$.
Figure~\ref{fig:NN_anyon} shows that increasing $V$ slows the propagation of the correlation front in exactly the way expected from the XXZ mapping.
The dashed guides illustrate the scaling $\Delta x\sim t^{1/z}$ with $z=1$ for $V=J$, $z=3/2$ at the isotropic point $V=2J$, and $z=2$ for $V=4J$ and $8J$.
The same sequence is visible in the on-site correlator shown in Fig.~\ref{fig:dens_dens_fit}, whose long-time decay follows
\begin{equation}
C^c_{00}(t) \sim t^{-1/z}.
\end{equation}
Thus the density-density correlations recover the standard XXZ hydrodynamics and remain insensitive to the anyonic statistical phase, in clear contrast to the Green's functions, where fractional statistics controls the chiral asymmetry.

\section{Conclusions}
\label{sec:conclusions}

In this work, we studied dynamical correlations of interacting hard-core anyons on a one-dimensional lattice at infinite temperature.
The main conclusion is that fractional statistics remains clearly visible in one-body dynamical observables even in a maximally mixed ensemble.
At $V=0$, the greater Green's functions are inversion symmetric for all statistical phases $\theta$, despite the presence of anyonic Jordan-Wigner strings.
The free dynamics interpolates smoothly between the two exactly solvable limits: a Gaussian, strongly localized bosonic correlator and an oscillatory fermionic correlator with ballistic spreading and $t^{-1/2}$ decay.
For intermediate statistical phases, the correlations retain the fermionic oscillation scale but acquire a faster, approximately exponential damping.

The central result concerns the effect of finite nearest-neighbor interactions.
Although the spectrum is independent of $\theta$, interactions generate a pronounced left-right asymmetry in the Green's functions for anyons with $0<\theta<\pi$.
This chirality originates from the noncommutativity of the interaction term with the nonlocal anyonic strings, and it is strongest at intermediate couplings $V\sim J$ where hopping and interactions compete most effectively.
At strong coupling, the dynamics crosses over toward an atomic-limit regime in which the dependence on $\theta$ is progressively suppressed.
In turn, the Green's function decay evolves toward a universal $t^{-1}$ behavior.
This same crossover is visible in the spectral observables: The free-limit density of states evolves from a bosonic Gaussian to fermionic Van Hove singularities, while for large $V$ both the local and momentum-resolved spectra approach a universal three-band structure centered near $\omega=0$ and $\omega=\pm V$.

The density-density correlations convey a complementary message.
Because they do not contain anyonic strings, they are independent of $\theta$ and recover the known infinite-temperature hydrodynamics of the XXZ chain: ballistic transport for $V<2J$, superdiffusion at $V=2J$, and diffusion for $V>2J$.

These results indicate a separation of the role of fractional statistics in one-body coherence from the transport of the conserved density.
They also identify dynamical correlation functions as a direct probe of anyonic statistics in high-entropy quantum systems.
Natural extensions include finite-temperature dynamics, multicomponent, or spinful anyons with on-site interactions, and the study of out-of-time-ordered correlators~\cite{Hashimoto2017,Swingle2018} as a further probe of interaction-induced asymmetry.

\begin{acknowledgments}
We acknowledge financial support from CNCS/CCCDI-UEFISCDI, under Projects No.~PN-IV-P1-PCE-2023-0987, and No.~PN-IV-P1-PCE-2023-0159, and by the ``Nucleu'' Program within the PNCDI 2022-2027, Romania, carried out with the support of MEC, Project No.~27N/03.01.2023, component project code PN 23 24 01 04, and also by ``Nucleu'' Grant No.~30N/2023.
Simulations were partly performed on resources made available by the computational
capacity of the Echidna HPC cluster, which is an integral part of the
Advanced computing platform for emerging technologies and related
fields - PACTE within INCDTIM Cluj-Napoca.
We acknowledge the University of Oradea for partially covering the publication costs of this article.
We also acknowledge the Digital Government Development and Project Management Ltd. for awarding us access to the Komondor HPC facility based in Hungary.
This work was also supported by the National Research, Development and Innovation Office - NKFIH  Project No.  K142179 and 
by the HUN-REN Hungarian Research Network through the Supported Research Groups
Programme, HUN-REN-BME-BCE Quantum Technology Research Group (TKCS-2024/34). 

\end{acknowledgments}
\section*{DATA AVAILABILITY}
The data that support the findings of this study are available from the corresponding author upon reasonable request.
\appendix

\begin{figure}[th]
\centering
\includegraphics[width=\columnwidth]{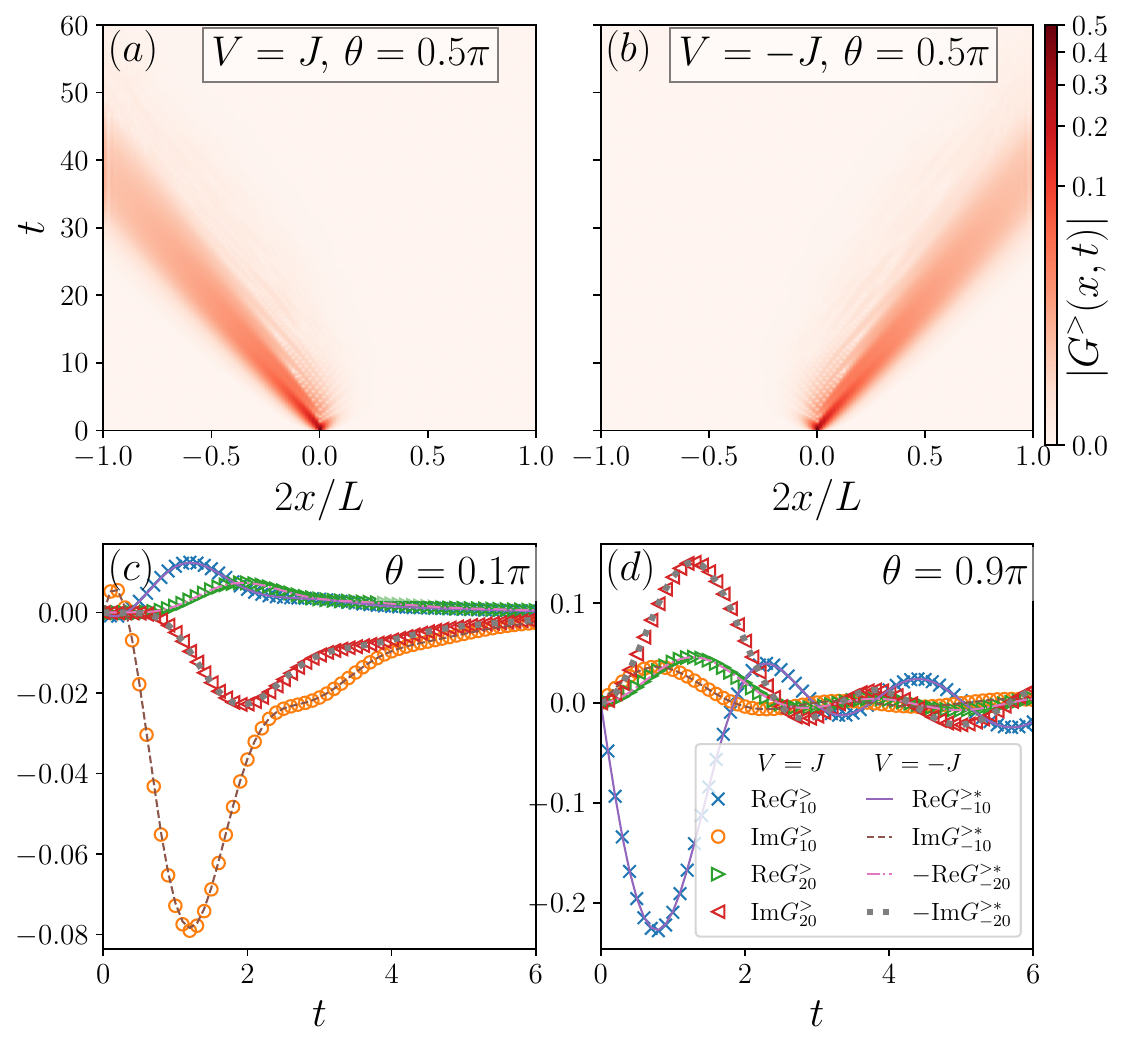}
\caption{The greater Green's function amplitude $|G^>(x,t)|$ at (a) $V$ and (b) $-V$ are related by spatial inversion symmetry.
Panels (c) and (d) illustrate a numerical verification of Eq.~\eqref{eq:GVs}, relating the greater Green's function at opposite interaction strengths, for $\theta=0.1\pi$ in panel (c) and $\theta=0.9\pi$ in panel (d) at $V=\pm J$, and $k=0$ and $j=\pm1$ and $\pm2$.
The legend in panel (d) applies also to panel (c) and $L=129$.}
\label{fig:G_x_t_sym} 
\end{figure}

\section{Symmetries of the correlation functions}
\label{app:symmetries}
Here, we illustrate how to derive the symmetry constraints on the Green's functions using the symmetry operators and then numerically verify the relation~\eqref{eq:GVs} between the Green's functions at opposite interaction strengths.
We also quantify the asymmetry for the Green's functions for several statistical phases and interaction strengths.

First, we focus on the spatial inversion symmetry operator $\mc I$ to determine Eq.~\eqref{eq:Ginv} for the greater Green's function.
Under inversion symmetry, a site $j$ is mapped to the site $j'=L-j+1$.
Let us consider an anyon operator $a_j(\theta)$ at site $j$ with explicit dependence on $\theta$,
\begin{align}\label{eq:ainv}
\mc I a_j(\theta) \mc T^\dag &= b_{j'}\mc I e^{i\theta\sum_{l<j} n_l} \mc I^\dag = b_{j'} e^{i\theta\sum_{l'>j'} n_{l'}} \notag\\
&= a_{j'}(-\theta) e^{-i\theta} e^{i\theta N},
\end{align}
where $N=\sum_l n_l$ is the total particle-number operator.
Inside the correlation function, we introduce the identity operator $\id=\mc I^\dag\mc I $,
\begin{align}
\avg{a_j(t)a_k^\dag}_\theta &= \avg{
e^{iHt}\mc I^\dag\mc I a_j(\theta)\mc I^\dag\mc I  e^{-iHt} \mc I^\dag\mc I a_k^\dag(\theta) \mc I^\dag \mc I
},
\end{align}
with the initial correlation function at statistical phase $\theta$, denoted in the subscript.
Applying Eq.~\eqref{eq:ainv} to the anyon operators, and using the invariance of the Hamiltonian under inversion symmetry, $\mc I H \mc I^\dag = H$, plus the commutation between $N$ and $H$, one obtains
\begin{equation}
\avg{a_j(t)a_k^\dag}_\theta = \avg{\mc I^\dag e^{iHt} a_{j'}(-\theta) e^{-iHt} a_{k'}^\dag(-\theta)\mc I}.
\end{equation}
Since the trace is invariant under unitary operators, it follows that
\begin{equation}
\avg{a_j(t)a_k^\dag}_\theta = \avg{a_{j'}(t) a_{k'}^\dag}_{-\theta},
\end{equation}
i.e., Eq.~\eqref{eq:Ginv}.
Similar procedures are performed to derive the other symmetry constraints on the Green's functions.

Finally, we use TEBD in Fock-Liouville space in order to verify the expression relating the Green's functions at opposite interaction strengths in a system of $L=129$ sites. 
Figures~\ref{fig:G_x_t_sym}(a) and~\ref{fig:G_x_t_sym}(b) exemplify the relation~\eqref{eq:GVs}, i.e., Green's functions at opposite interaction strengths $V=\pm J$ at some arbitrary statistical phase (here $\theta=0.5\pi$) are related by inversion symmetry.
Further analysis in Fig.~\ref{fig:G_x_t_sym}(c) and Fig.~\ref{fig:G_x_t_sym}(d) also verifies the even-odd effects in the Green's-function phase of Eq.~\eqref{eq:GVs} for $G_{01}$ and $G_{02}$ at the representative phases $\theta=0.1\pi$ and $\theta=0.9\pi$.
The results have been also confirmed by using ED simulations in the full Fock space for small systems $L\in\{9,11\}$.

\begin{figure*}[ht]
\centering
\includegraphics[width=1.0\textwidth]{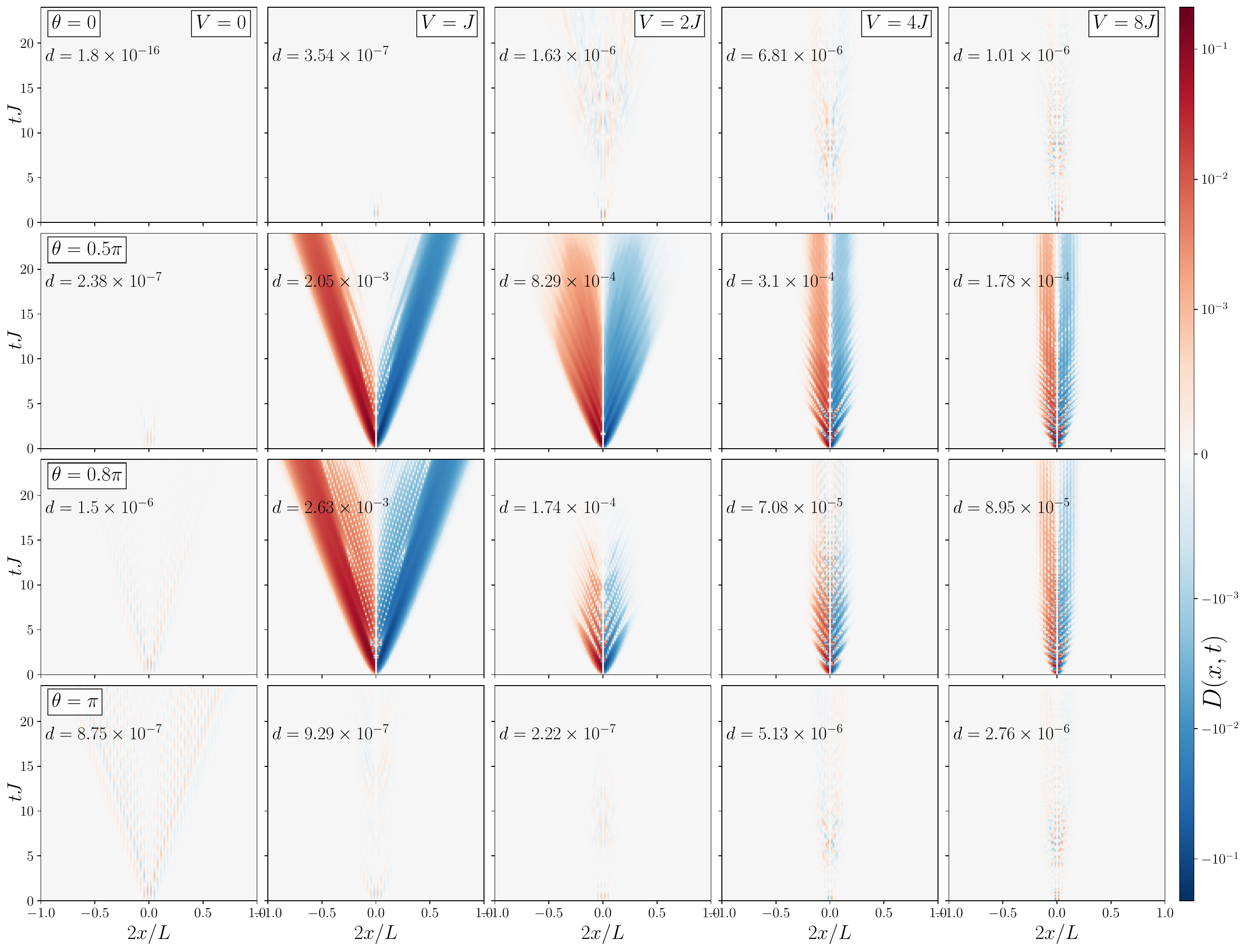}
\caption{
    The inversion asymmetry is represented by plotting the space antisymmetric function $D(x,t) = |G^>(x,t)|-|G^>(-x,t)|$ for several statistical phases (rows) and interaction strengths (columns) as in Fig.~\ref{fig:G_x_t_Vs}. The asymmetry measure $d(L,\tau)$~\eqref{eq:dLT} is annotated in each panel.
    The scale is linear up to $10^{-3}$ and logarithmic beyond.
}
\label{fig:asymmetry}
\end{figure*}

Finally, in order to further investigate and better represent the inversion asymmetry in the Green's functions, we define the space antisymmetric function
\begin{equation}
D(x,t) = |G^>(x,t)|-|G^>(-x,t)|.
\end{equation}
For our lattice simulations, we may assign a single number $d(L,\tau)$ to quantify the asymmetry of the Green's functions by averaging $D(x,t)$ over space and time,
\begin{equation}
\label{eq:dLT}
d(L,\tau) = \frac{2}{L\tau}\int_{0}^{L/2} dx'\int_0^\tau dt'\, |D(x',t')|,
\end{equation}
with $L$ the system size and $\tau$ the total simulation time.
The function is expected to vanish for $V=0$ and for $\theta=0$ and $\theta=\pi$, while it is nonzero for anyons with $0<\theta<\pi$ and $V\neq 0$.
Figure~\ref{fig:asymmetry} shows the time evolution of $D(x,t)$ for several statistical phases and interaction strengths, illustrating that the asymmetry is strongest at intermediate interaction strengths $V\sim J$ for generic anyons away from the bosonic and fermionic limits.
The associated asymmetry measure $d(L,\tau)$ is annotated on the figures.
For $d(L,\tau)\sim \mc O(10^{-6})$, as seen for fermions and in other cases from Fig.~\ref{fig:asymmetry}, data are unreliable since they corresponds to numerical noise accumulated in the TEBD simulations, due to finite bond dimension $\chi=128$ and a time step $\delta t=0.1J$.
In contrast, the anyons with $V=J$ show an asymmetry measure $d(L,\tau)\sim \mc O(10^{-3})$, which is several orders of magnitude larger than the numerical noise.

\section{Methods}
\label{app:methods}

To simulate the dynamics of the correlation functions in the anyon system, we use several numerical techniques.
For the noninteracting case $V=0$, we use methods based on Refs.~\cite{Patu2015,Wang2022} in order to obtain high-quality data, correct to machine precision, that describe the correlation functions in large $\sim \mc O(10^2)$-site systems.

Exact diagonalization in the presence of interactions remains possible only for small systems $\sim \mc O(10)$ sites due to the exponential growth of the Hilbert space for many-body systems.
Thus, we use two TEBD methods to study the time evolution of correlation functions. 
Their results are benchmarked against exact diagonalization for small systems.

\subsection{Exact diagonalization}\label{sec:ED}
At $V=0$, the anyon correlation functions may be computed using exact diagonalization by performing algebraic manipulations of matrices of size $L\times L$.
We follow and summarize the method developed in Ref.~\cite{Wang2022}.

The free-anyon Hamiltonian is quadratic:
\begin{equation}
H = \bm c^\dag \mc H \bm c,\quad \bm c^\dag = (c_1^\dag,c_2^\dag,\ldots,c_L^\dag),
\end{equation}
with single-particle Hamiltonian matrix $\mc H$ given by
\begin{equation}
[\mc H]_{kl} = J(\delta_{k,l+1}+\delta_{k+1,l}).
\end{equation}
The anyon operators may be represented in terms of fermion operators~\eqref{eq:JW} as exponentials of quadratic forms,
\begin{equation}
a_j = c_j e^{-\bm c^\dag T^{(j)}\bm c}.
\end{equation}
The matrices $T^{(j)}$ are diagonal with elements equal to $i(\pi-\theta)$ for all sites left of site $j$, and zero otherwise, i.e.,
\begin{equation}
[T^{(j)}]_{kl} = i\theta'\delta_{kl} \Theta(j-k-1),\quad \theta' = \pi-\theta,
\end{equation}
with the Heaviside step function $\Theta(x)=0$ for $x<0$ and $\Theta(x)=1$ for $x\geq 0$.

In order to compute the anyon correlation functions, the fermion operators are commuted through the exponentials of quadratic forms such as the Hamiltonian and the $T$ matrices in order to express the correlations as Gaussian fermionic traces,
\begin{align}
Z_M&=\tr[e^{\bm c^\dag M \bm c}] = \det(1+e^M),\notag\\    
\avg{c_jc_k^\dag}_M&=\frac1{Z_M}\tr[c_jc_k^\dag e^{\bm c^\dag M \bm c}] = [(1+e^M)^{-1}]_{jk},\notag\\
\avg{c_jc_kc_l^\dag c^\dag_m}_M&=
\avg{c_jc_m^\dag}_M \avg{c_kc_l^\dag}_M - \avg{c_jc_l^\dag}_M \avg{c_kc_m^\dag}_M.
\end{align}
The commutations make use of the known relations for exponentials of quadratic forms,
\begin{equation}
e^{\bm c^\dag A \bm c} c_j e^{-\bm c^\dag A \bm c} = [e^{-A}]_{jk} c_k,
\end{equation}
and\begin{equation}
e^{\bm c^\dag A \bm c} e^{\bm c^\dag B \bm c} 
=  e^{\bm c^\dag D \bm c},\quad e^D = e^A e^B, 
\end{equation}
with $A$, $B$, and $D$ being arbitrary square matrices of equal size $L\times L$.
After such a series of algebraic manipulations, one arrives at an expression for the dynamical correlations in the greater Green's function,
\begin{equation}
\avg{a_j(t)a_k^\dag}
=
\frac{\det(1+O^{jk})}{2^L}
\left[e^{-i\mc Ht}
\frac{1}{1+(O^{jk})^{-1}}
\right]_{jk},
\end{equation}
with
\begin{equation}
O^{jk} = e^{i\mc H t} e^{-T^{(j)}} e^{-i\mc H t} e^{T^{(k)}}.
\end{equation}
The Green's function is now evaluated efficiently by calculating only matrix multiplications, diagonalizations, and determinants of $L\times L$ matrices, for systems of hundreds of sites, $L\sim \mc O(10^2)$.

At finite interactions $V\neq 0$, we use an exact representation of operators in the full many-body Fock space.
The Hamiltonian is completely diagonalized using standard linear algebra libraries, which in turn allows us to represent exactly the time-evolution operator.
Then, the anyon operators are time evolved in the Heisenberg picture and correlations are computed at arbitrary times.
These exact techniques are limited by the exponential growth of the Hilbert space and are therefore restricted to small system sizes $L\sim \mc O(10)$.
These simulations are used to benchmark and verify the accuracy of the other methods for small systems.

\begin{figure}[th]
\includegraphics[width=\columnwidth]{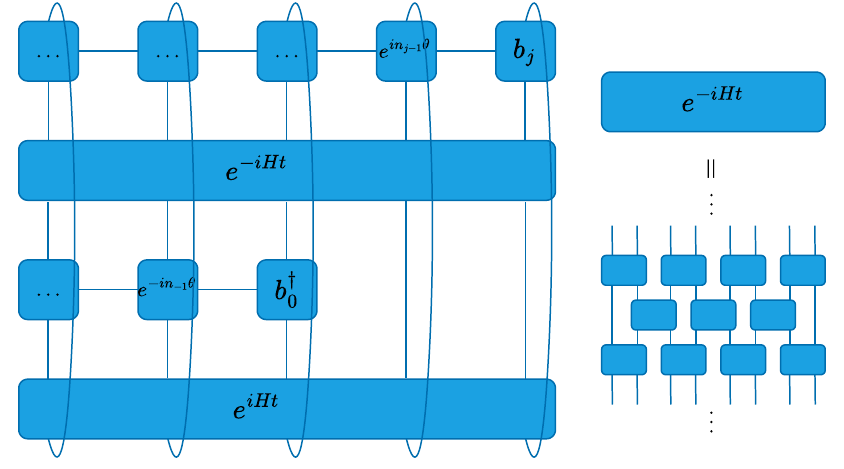}
\caption{Tensor-network representation of the correlation function $\avg{a_j(t)a_0^\dag}$ at infinite temperature. 
The anyons are represented as an MPO of hard-core bosons with Jordan-Wigner strings that depend on angle $\theta$.
The evolution is performed using a second-order Suzuki-Trotter decomposition with a brick-wall pattern.}
\label{fig:schematic1}
\end{figure}

\subsection{Operator evolution using TEBD}
For larger systems, we use time-evolving block decimation techniques based on matrix product states~\cite{Schollwoeck2011,Fishman2022}.

At infinite temperature, the density matrix is a tensor product of identity operators, 
\begin{equation}
\rho_\infty = \frac{1}{2^L}\bigotimes_{j=-L/2}^{L/2}\id.    
\end{equation}
The anyon operators are represented in simulations as matrix product operators (MPOs) of hard-core bosons with $\theta$-dependent Jordan-Wigner strings~\eqref{eq:JW},
\begin{equation}
a_j = \bigotimes_{l<j} e^{i\theta n_l} \otimes b_j \bigotimes_{l>j} \id.
\end{equation}

The correlations in the greater Green's function are computed by setting the reference site $k$ at the center of the chain, $k=0$. 
Since the density matrix is proportional to the identity operator, one can also write the correlation function as
\begin{equation}
\avg{a_j(t) a_0^\dag} = \avg{a_j a_0^\dag(-t)},
\end{equation}
where $a_0$ is evolved backward in time.
This choice is more convenient in TEBD simulations, since a single anyon MPO is time evolved, instead of evolving multiple anyon MPOs for each site $j$.
The schematic representation of the calculation of the dynamical correlations is shown in Fig.~\ref{fig:schematic1}.
The time-evolution operator is implemented using a second-order Suzuki-Trotter decomposition with a brick-wall pattern of gates.
The bond dimension in simulations is typically $\chi=128$.

\subsection{TEBD in Fock-Liouville space}
To compute the correlation functions at finite interactions, we also use an alternative method based on the vectorization of the density matrix in Fock-Liouville space~\cite{Dzhioev2011}.

This approach is motivated by the idea of mapping the von Neumann equation for the density matrix $\rho(t)$ to a Schr\"odinger-like equation for a state vector in a doubled Fock space:
\begin{equation}
\frac{d}{dt}\rho(t) = -i[H,\rho(t)].
\end{equation}
The density matrix is vectorized under the Choi-Jamio\l{}kowski (CJ) isomorphism~\cite{Choi1972,Jamiolkowski1972} as 
\begin{equation}
\rho = \sum_{m,n} \rho_{mn} |m\rangle\langle n| \mapsto |\rho \rrangle = \sum_{m,n} \rho_{mn} |m\rangle \otimes |n\rangle,
\end{equation}
and thus the density matrix becomes a vector in a doubled Fock space, also called Fock-Liouville space.
Action of an operator on the density matrix is mapped under the CJ isomorphism as
\begin{equation}
A\rho B \mapsto A\otimes B^T |\rho \rrangle,
\end{equation}
with $A$ and $B$ being some arbitrary operators.
Traces over operators become inner products of vectors under the CJ isomorphism,
\begin{equation}
\tr[A^\dag B] \mapsto \llangle A|B\rrangle,
\end{equation}
with $\llangle A| = |A\rrangle^\dag$.
Thus, the von Neumann equation is mapped to a Schr\"odinger-like equation,
\begin{equation}
\frac{d}{dt}|\rho(t)\rrangle = -i\mc L |\rho(t)\rrangle,
\end{equation}
where the Hamiltonian's role is taken by the Liouville (super)operator,
\begin{equation}\label{eq:liouv}
\mc L = H\otimes \id - \id \otimes H^T.
\end{equation}

This mapping allows us to use TEBD techniques on states living in the Fock-Liouville space, which are represented as MPSs.
The correlation functions are computed as follows:
\begin{equation}\label{eq:inner}
\tr[a^\dag_j(t)a_0] = \tr[a_j^\dag e^{-iHt} a_0 e^{iHt}] = \llangle a_j| e^{-iH t}\otimes e^{iH^T t} |a_0\rrangle.
\end{equation}
One recognizes in the above expression a time-evolved state vector $|a_0(t)\rrangle$ with the Liouville operator~\eqref{eq:liouv},
\begin{equation}
|a_0(t)\rrangle = e^{-i\mc L t} |a_0\rrangle.
\end{equation}
In simulations, we use the MPS representation of the state vector $|a_j\rrangle$.
The anyonic state is represented as a hard-core boson with anyonic Jordan-Wigner strings,
\begin{equation}
|a_j\rrangle = e^{i\theta\sum_{k<j} n_k\otimes \id} (b_j\otimes \id) |\id\rrangle.
\end{equation}
Then, the time evolution is performed using the Liouville operator $\mc L$.
As before, the time-evolution operator is implemented using a second-order Suzuki-Trotter decomposition with a brick-wall pattern of gates and, usually, the bond dimension in calculations is $\chi=128$.
The final correlation functions are computed by taking the inner product in Fock-Liouville space~\eqref{eq:inner}.

\bibliography{references}

\end{document}